\documentclass[12pt]{iopart}

\usepackage[usenames,dvipsnames]{xcolor}
\usepackage{graphicx}
\usepackage{iopams}
\usepackage{subeqn}
\usepackage{cite}
\usepackage{comment}

\newcommand{\bdm}{\begin{displaymath}}
\newcommand{\edm}{\end{displaymath}}

\newcommand{\be}{\begin{equation}}
\newcommand{\ee}{\end{equation}}

\renewcommand{\bi}{\begin{itemize}}
\newcommand{\ei}{\end{itemize}}

\newcommand{\mean}[1]{\left\langle #1 \right\rangle}

\newcommand{\RDM}{\rho}
\newcommand{\ket}[1]{\left|#1\right\rangle}

\renewcommand{\imath}{\mathrm{i}}

\begin{document}

\title{Multi-phonon relaxation and generation of quantum states
  in a nonlinear mechanical oscillator}

\author{A. Voje$^*$, A. Croy and A. Isacsson}

\address{Department of Applied Physics, Chalmers University of
  Technology,
  SE-412 96 G{\"o}teborg, Sweden\\
  $^*$Corresponding author: aurora@chalmers.se
}
\date{Version: \today}
\begin{abstract}
The dissipative quantum dynamics of an anharmonic oscillator is
investigated theoretically in the context of carbon-based
nano-mechanical systems. In the short-time limit, it is known that
macroscopic superposition states appear for such oscillators. In the
long-time limit, single and multi-phonon dissipation lead to
decoherence of the non-classical states. However, at zero temperature,
as a result of two-phonon losses the quantum oscillator eventually
evolves into a non-classical steady state. The relaxation of this
state due to thermal excitations and one-phonon losses is numerically
and analytically studied. The possibility of verifying the occurrence
of the non-classical state is investigated and signatures of the
quantum features arising in a ring-down setup are presented. The
feasibility of the verification scheme is discussed in the context of
quantum nano-mechanical systems.
\end{abstract}
\submitto{\NJP}


\maketitle 

%
\section{Introduction} 
Nanoelectro-mechanical (NEM) resonators such as cantilevers, membranes
and beams are important systems for the study of quantum phenomena in
macroscopic, mechanical man-made objects \cite{poza12}. For about a
decade, the goal has been to cool the NEM resonators to the ground
state and to verify the achievement of this state. This has recently
been accomplished\cite{ocho+10,tedo+11,chal+11}, and the next step is
the generation, detection and coherent manipulation of more intricate
quantum states, e.g., superpositions of macroscopic states
(Schr{\"o}dinger cat states) \cite{voki+12} and Fock states \cite{riki+12}.

While a harmonic quantum oscillator displays a behavior analogous to
its classical counterpart, the same does not hold for nonlinear
oscillators. The presence of a conservative (Kerr) nonlinearity
facilitates the creation of non-classical states. Although top-down
fabricated NEM-resonators typically possess a finite Kerr constant
$\mu$, it is usually too small to alone suffice for non-classical
state generation. In contrast, carbon nanotubes (CNTs) and graphene
membranes can exhibit strong enough nonlinearities for non-classical
states to emerge during free evolution\cite{atis+08, voki+12}. Using
CNTs or graphene also alleviates the need for active cooling due to
their low masses and high frequencies\cite{poza12,voki+12}. For this
reason, carbon based NEM-resonators are highly promising for the study
of nonlinear mechanical systems in the quantum regime.

If a system possesses conservative nonlinearities, it is natural to
also take into account higher orders in the system-environment
coupling. This results in nonlinear damping
(NLD)\cite{zw73,lise81,dykr84,zash+12}. In carbon based NEM-resonators
it has recently been shown that the dissipative mechanisms can indeed
be nonlinear in the oscillator coordinate\cite{eimo+11, crmi+12}.
Hence, to fully understand the quantum dynamics of carbon based NEM
resonators it is essential to comprehend the interplay between
conservative nonlinearities and NLD. In general, NLD will create
interesting quantum features rather than remove
them\cite{silo78,gikn93,giga+94,lo84,evsp+12}. For example, as is well
known in quantum optics\cite{sh67,giga+94,silo78,gikn93}, for a
harmonic oscillator, the steady state $\rho$ obtained from
NLD due to two-photon losses at zero temperature ($T=0$) is a mixed
state with $\rm{Tr}[\rho^2]<1$. Moreover, because of the parity
conserving nature of the NLD, this steady state density matrix
has coherences in terms of nonzero off-diagonal elements $\rho_{01}$
in the number basis. Recently, it was proposed to use this effect in
order to generate and preserve Schr\"{o}dinger cat states in a double
well potential realized by a SQUID ring\cite{evsp+12}. In an
optomechanical setup non-thermal and squeezed states were shown to be
obtainable by two-phonon cooling \cite{nubo+10}.

In this paper we study a situation more relevant to the coherent
quantum dynamics of carbon based NEM-resonators, namely, a nonlinear
oscillator with NLD. In this field there is little to be found in the
literature{\cite{dykr84}}. Our main goals are to characterize the
resulting steady states along with the corresponding features of
the relaxation dynamics. For the nonlinear oscillator with NLD we find
that the parity conserving nature of NLD helps to preserve phase
coherence also in the presence of a finite conservative
nonlinearity. However, we also find that for sufficiently large ratio
between the Kerr constant $\mu$ and the NLD strength $\gamma_2$, the
coherence of the steady state is suppressed. Further, at finite
temperatures, with NLD, the final state is not a standard thermal
state with a reduced density matrix $\rho\propto e^{-H/k_BT}$ unless
linear damping (LD) is present. The corresponding Wigner distribution
has a dip in the center, whose depth depends on the initial
conditions. In the presence of both NLD and LD, the relaxation is
governed by a crossover between two different relaxation rates. In the
long-time limit, the decay rate $\Gamma$ is independent of the initial
oscillator amplitude for NLD. Instead it is a function only of
temperature $T$. To facilitate experimental determination of the
governing damping mechanism, we also present the corresponding
signatures in the time evolution of the position coordinate variance.

This paper is organized as follows: In section \ref{sec:model}, we
present the underlying theoretical frameworks used for analytical
(rotating wave approximation [RWA]) and numerical
(Wangsness-Bloch-Redfield\cite{wabl53,bl57,re65}) calculations. The
latter are used to verify the validity of the RWA-calculations. Then,
in section \ref{sec:stationary}, we present results pertaining to the
steady states for various scenarios involving NLD and LD. Results
on the relaxation towards the steady states are found in section
\ref{sec:relaxation}, followed by a section with summary and conclusions.

\section{Model and Methods}\label{sec:model}
\subsection{Model}
To model the dynamics of the anharmonic oscillator subject to linear
and nonlinear damping, we consider a single mechanical mode with
frequency $\Omega$ and Kerr constant $\mu$. The mode is coupled
linearly and nonlinearly to a reservoir of harmonic oscillators with
mode frequencies $\omega_k$. The mode mass and the reduced Planck
constant are set to be $m=1$ and $\hbar=1$,
respectively. Correspondingly, the total Hamiltonian is
\begin{subeqnarray}\label{eq:hamiltonian_terms}
H & =& H_{\rm S}+ H_{\rm B}+ H_{\rm
  SB},\label{eq:hamiltonian_terms1}\\ H_{\rm S} & = &
\frac{1}{2}p^2+\frac{1}{2}\Omega^2 q^2 + \frac{2\mu}{3}q^4
,\label{eq:hamiltonian_terms2}\\ H_{\rm B} & = &\sum_k \omega_k
b^{\dag}_k b_k,\label{eq:hamiltonian_terms3}\\ H_{\rm SB}& = &q
\sum_k(\sqrt{2}\eta_{1 k})(b^{\dag}_k + b_k) + q^2 \sum_k(2 \eta_{2
  k}) (b^{\dag}_k + b_k),\label{eq:hamiltonian_terms4}
\end{subeqnarray}
where the mode position and momentum operators are $q=q_0(a^{\dag}
+a)/\sqrt{2}$, $p=\imath (a^{\dag} -a)/\sqrt{2}q_0$, and similarly the
bath operators are given by $x_k=x_{0k}(b_k^{\dag} +b_k)/\sqrt{2}$,
$p_k=\imath (b_k^{\dag} -b_k)/\sqrt{2}x_{0k}$. Here $q_0 = \sqrt{1/
  \Omega}$ and $x_{0k} = \sqrt{1/m_k \omega_k}$ are the zero point
amplitudes of the oscillator and the bath, and satisfy the canonical
commutation relations. The linear and the nonlinear coupling constants
of the $k$'th reservoir mode to the resonator mode are denoted by
$\eta_{1 k}$ and $\eta_{2 k}$, respectively. These couplings lead to
amplitude-independent (linear) and amplitude-dependent (non-linear)
\cite{zw73,lise81,dykr84,zash+12} damping of the oscillator
motion. The former coupling involves the exchange of single vibration
quanta (vibrons) with the reservoir, while in the latter case two
vibrons are transferred simultaneously.

We consider the following scenario: The oscillator mode is assumed to
be initially in the ground state $\ket{0}$. Then, by applying a
voltage pulse for a finite time, the oscillator is displaced. The
corresponding state is, by definition \cite{gl63}, a coherent state,
$\Psi(t=0)=D(\alpha)\ket{0}\equiv\ket{\alpha}$. Here,
$D(\alpha)=\exp(\alpha a^\dagger - \alpha^* a)$ is the displacement
operator and the amplitude $\alpha$ depends on the strength and
duration of the voltage pulse. During dissipation-free evolution
under $H_{\rm S}$ only, a coherent state will evolve into a
Yurke-Stoler cat state $(\vert \alpha\rangle + \rm i \vert-
\alpha\rangle)/\sqrt{2}$. This state will periodically re-appear with
a frequency set by the Kerr constant $\mu$. Typically the state's
emergence is slow compared to the system's eigenfrequency $\Omega$
\cite{yust86}. When taking into account the interaction with the
reservoir, the created cat states will be influenced by the
dissipation and the oscillator will eventually relax into a steady
state. The state of the mechanical mode during the relaxation is
described by the reduced density matrix (RDM) $\RDM=\tr_{\rm B}
\rho_{\rm T}$, which is obtained from the state of the total system
$\rho_{\rm T}$ by tracing over bath degrees of freedom.
 
The time-evolution of the reduced density matrix is described by a
(generalized) quantum master equation (QME) \cite{brpe02}. Here we
follow the standard approach using the Born-Markov approximation to
obtain the QME from the von Neumann equation for the total density
matrix \cite{brpe02}. In this approximation the QME in the interaction
picture with respect to $H_{\rm S}$ is given by
\begin{eqnarray}\label{eq:QMEwIntfinal}
\frac{\partial }{\partial t}\RDM(t) =& 
-\sum_{l,m}\int_0^{\infty} d\tau\Big[ 
A_l(t)A_m(t-\tau)\RDM(t)
-
A_m(t-\tau)\RDM(t) A_l(t) 
\Big]
C_{l m}(\tau) \nonumber\\
&\:+
\Big[\:
\RDM(t) A_m(t-\tau)A_l(t)
-
A_l(t)\RDM(t) A_m(t-\tau)
\Big]
C_{m l}(-\tau).
\end{eqnarray}
The operators $A_m$ are given by the expansion of the interaction
Hamiltonian in the interaction picture
\be\label{eq:genInt}
	H_{SB}= \sum_{m=1,2} A_m(t)\otimes B_m(t)\;, 
\ee
where
\begin{equation}
	A_m(t) = e^{\imath H_{\rm S} t}\;(a^{\dag} + a)^m \;e^{-\imath
          H_{\rm S} t},\quad B_m(t)= \sum_k {q_0^m \eta_{m
            k}}\left(b^{\dag}_k e^{\imath \omega_k t} + b_k e^{-\imath
          \omega_k t}\right).
\end{equation}
Assuming the reservoir to be initially in thermal equilibrium, the
reservoir correlation functions are given by
\begin{eqnarray}\label{eq:corrfunct}
	C_{m l}(\tau)&=& \tr_B \{B_m(t) B_l(t-\tau)\rho_{\rm B} \}
        \nonumber\\ &=& \int\frac{d\omega}{2\pi}\kappa_{m l}(\omega)
        \left[ n_{\rm B}(\omega) e^{\imath \omega \tau} + (n_{\rm
            B}(\omega) + 1) e^{-\imath \omega \tau} \right]\;,
\end{eqnarray}
where $n_{\rm B}(\omega)=[e^{\omega/k_{\rm B}T}-1]^{-1}$ is the Bose
distribution function, $\rho_{\rm B}$ is the thermal state and
$\kappa_{m l}(\omega)=\kappa_{l m}(\omega)=2\pi \sum_k (\eta_{m
  k}\eta_{l k} q_0^{m+l}) \delta(\omega-\omega_k)$ is the spectral
density of the environment coupling.

In general, the frequency dependence of $\kappa_{m l}$ is determined
by the specific geometry of the system.  As long as the spectral
density is smooth and slowly varying around $\omega=\Omega$ and
$\omega=2\Omega$ the detailed dependence on $\omega$ is not
important\cite{dy12} (see also \ref{sec:RWA}). To be specific,
we consider an Ohmic spectral density \cite{brpe02}, i.e., we take
\be 
\kappa_{m l}(\omega) = \Gamma_{m l} \frac{\omega}{\Omega\sqrt{m
    l}}\;.  
\ee
The values of $\Gamma_{1 1}$ and $\Gamma_{2 2}$ are used to specify
the strength of LD and NLD, respectively. The
off-diagonal value $\Gamma_{1 2}$ can, in general, be chosen
independently of $\Gamma_{1 1}$ and $\Gamma_{2 2}$. If the oscillator
couples to two independent baths, the off-diagonal spectral density
vanishes.  In the RWA, which is discussed in \ref{sec:RWA}, the
dissipation terms associated with $\Gamma_{1 2}$ can be neglected.
Here we set $\Gamma_{1 2} = \sqrt{\Gamma_{1 1} \Gamma_{2 2}}$, which
vanishes if either linear or nonlinear damping is absent.

\subsection{Relaxation dynamics in the rotating frame}\label{sec:RWA}
Typically, for NEMS the time scales associated with $\Omega$ and $\mu$
are well separated with $\Omega\gg\mu$ (see also section
\ref{sec:summary}), which justifies an analytic treatment in the
RWA. Within this approximation the QME (\ref{eq:QMEwIntfinal}) in the
Schr\"{o}dinger representation becomes\cite{gazo04}
\be\label{eq:gen_lindblad} \partial_t \rho(t)= (\mathcal{L}_0 +
\mathcal{L}_1 + \mathcal{L}_2)\rho(t), \ee
where the Lindblad superoperators are given by
\begin{eqnarray}
\begin{array}{lll}
\mathcal{L}_0 \rho & = & -i(\Omega+\mu)[a^{\dag}a,\rho]
-i\mu[(a^{\dag}a)^2,\rho], \\ \mathcal{L}_1 \rho & = & \gamma_{-} a
\rho a^{\dag} -\frac{\gamma_{-}}{2}\{a^{\dag}a,\rho \} + \gamma_{+}
a^{\dag} \rho a -\frac{\gamma_{+}}{2}\{aa^{\dag},\rho \},
\\ \mathcal{L}_2 \rho& = & \gamma_{2-} aa \rho a^{\dag}a^{\dag}
-\frac{\gamma_{2-}}{2}\{a^{\dag}a^{\dag}aa,\rho \} +
\gamma_{2+}a^{\dag} a^{\dag} \rho aa
-\frac{\gamma_{2+}}{2}\{aaa^{\dag}a^{\dag},\rho \}.
\end{array}
\end{eqnarray}
The superoperators $\mathcal{L}_0$, $\mathcal{L}_1$ and
$\mathcal{L}_2$ describe the coherent evolution due to $H_{\rm S}$,
linear and the nonlinear interaction with the reservoir,
respectively. The linear and nonlinear dissipation rates, $\gamma_\pm$
and $\gamma_{2\pm}$, are Fourier transforms of the correlation
function (\ref{eq:corrfunct}), and are given by
\begin{eqnarray}
\begin{array}{ll}
\gamma_{+} =\Gamma_{11}n_{\rm B}(\Omega), &
\gamma_{2+}=\Gamma_{22}n_{\rm B}(2\Omega),\\ 
\gamma_{-}=\Gamma_{11}[n_{\rm B}(\Omega)+1],&
\gamma_{2-}=\Gamma_{22}[n_{\rm B}(2\Omega)+1]. 
\end{array}
\end{eqnarray}
We would like to note, that the RWA equations given above are only
valid for a weak nonlinearity\cite{gazo04}. This is reflected by the
dependence of the rates on the transition frequency of the harmonic
part of the Hamiltonian $H_{\rm S}$ and not on the frequencies of the
full nonlinear $H_{\rm S}$. The validity of this approximation in the
present case will be shown in the next sections by comparing to
numerical calculations which include the exact transition frequencies
(see also section \ref{sec:numcalc}).  

The solution to \eref{eq:gen_lindblad} is obtained by rewriting the
matrix equation as a set of decoupled equations. To this end, $\RDM$
is projected onto the number basis $\RDM_{n, n + \lambda} = \langle n
\vert \rho \vert n + \lambda \rangle$, where $\lambda =
0,1,2,\ldots$. It is convenient to define $\RDM_{n, n+
  \lambda}(t)\equiv\sqrt{\frac{n!}{(n+\lambda)!}}  \psi_n(\lambda,t)
e^{\imath \lambda(\Omega+\mu) t}$ \cite{silo78}, which also removes
oscillations due to the term linear in $a^\dag a$ in
\eref{eq:gen_lindblad}. For fixed $\lambda$ the function
$\psi_n(\lambda,t)$ gives the entries of the $\lambda$'th super
diagonal of $\RDM$. Hence, \eref{eq:gen_lindblad} becomes
\begin{eqnarray}\label{eq:nonlinQMEPsi}
\partial_t \psi_n(\lambda,t) & = &
\gamma_{2+}(n+\lambda)(n+\lambda-1)\psi_{n-2} + \gamma_{2-}
(n+2)(n+1)\psi_{n+2} \nonumber\\
& &- \{-\imath \mu \lambda(\lambda + 2n) 
+ \gamma_{2-} n(n-1+\lambda) \nonumber\\
& & +\frac{\gamma_{2-}}{2} \lambda(\lambda-1) + \gamma_{2+}(n+1)(n+2+\lambda) 
+\frac{\gamma_{2+}}{2} \lambda(\lambda+1)\}
\psi_n \nonumber\\
& & + \gamma_{-} (n+1) \psi_{n+1} + \gamma_{+} (n+\lambda) \psi_{n-1} \nonumber\\
& & - \gamma_{-} 
  (n + \frac{1}{2}\lambda)\psi_n - \gamma_{+} 
  (n + 1+ \frac{1}{2}\lambda)\psi_n .
\end{eqnarray}
As indicated above, the equations for different $\lambda$ do not
couple. In particular, for $\lambda=0$, we obtain a single equation
for the populations $\RDM_{n,n}(t)=\psi_n(0,t)$, which is used in the
following to analyze the relaxation to the steady state.

In general, when all coefficients in \eref{eq:nonlinQMEPsi} are
non-zero the time evolution of $\psi_n(\lambda,t)$ has to be found
numerically. If either $\gamma_-$ or $\gamma_{2-}$ is vanishing, i.e.,
there are only one or two vibron losses present, the time-dependence
of $\psi_n(\lambda,t)$ can be calculated analytically (see
\cite{miho86,ph90,mo06} and \cite{silo78}). In these cases it is also
straight-forward to find the steady state solutions for all
$\lambda$. An exact solution can also be obtained in terms of the
  positive P function where the nonlinear damping is equivalent to the
  real part of a complex Kerr constant \cite{drwa79}. At finite
temperatures, when in addition to the losses, one or two vibron
excitations ($\gamma_\pm > 0$ or $\gamma_{2\pm} > 0$) contribute, the
stationary (time independent) probability distribution
$\RDM_{n,n}$ can be found using the detailed balance condition
\cite{la62, ha74} (see \ref{sec:app_fin_temp}). If one and two-vibron
processes compete ($\gamma_\pm > 0$ and $\gamma_{2\pm} > 0$), a
solution for $\RDM_{n,n}$ can be given in terms of a continued
fraction \cite{haha79} or (for $\gamma_{2+}=0$) in terms of confluent
hypergeometric functions\cite{domi97}.

In the next sections we are in particular interested in the behavior
of the position variance
\begin{equation}\label{eq:quad_def}
	\mean{ \Delta q^2 } =  \mean{ q^2 } - \mean{ q }^2\;,
\end{equation}
where $\mean{\bullet} = \tr_{\rm S} [ \bullet \RDM]$. Expressing $q$
in terms of creation and annihilation operators in the rotating frame
one finds
\begin{equation}\label{eq:quad_def_rwa}
	\mean{ \Delta q^2 }/q_0^2 \approx \frac{1}{2}\left[ \mean{
            a^\dagger a + a a^\dagger } - 2 |\mean{ a^\dagger }|^2
          \right] =\frac{1}{2} + \mean{ a^\dagger a } - |\mean{
          a^\dagger }|^2\;,
\end{equation}
where all oscillating contributions have been discarded.  The
expectation values of $a^\dagger a$ and $a^\dagger$ can be expressed
in terms of $\psi_{n}(0,t)$ and $\psi_{n}(1,t)$, respectively,
\begin{eqnarray}
	\mean{ a^\dagger a} 
	&=& \sum^\infty_{n=0}  n\, \RDM_{n, n}
	= \sum^\infty_{n=0} n\,\psi_{n}(0,t)\;, \\
	\mean{ a^\dagger } 
	&=& \sum^\infty_{n=0}  \sqrt{n+1} \RDM_{n, n+1}
	= \sum^\infty_{n=0} \psi_{n}(1,t)\;.
\end{eqnarray}
Knowing $\psi_{n}(0,t)$ and $\psi_{n}(1,t)$ therefore allows for an
evaluation of the position variance. In the following we focus on
obtaining such solutions in the long-time limit.

\subsection{Numerical Computation}\label{sec:numcalc}
In order to investigate the oscillator dynamics and to verify the
results obtained by the RWA approximation, the QME
\eref{eq:QMEwIntfinal} is solved numerically by
Wangsness-Bloch-Redfield \cite{wabl53,bl57,re65} calculations.  To
this end the operators $A_m$ are represented in the eigenbasis of the
oscillator Hamiltonian $H_{\rm S}$ and the Hilbert space is truncated
after $30$ states ($40$ states for $\alpha=3$), which yields converged
results. The initial coherent state, $\RDM(0)=\vert \alpha \rangle
\langle \alpha \vert$, is generated by applying the displacement
operator $D(\alpha)$ to the ground state.

%
%
\section{Steady state}\label{sec:stationary}

\subsection{Zero temperature}\label{sec:statzeroT}
We begin with the RWA-analysis of the steady state $\rho(\infty)$
at zero temperature. To evaluate the applicability of the RWA, we
compare the results to numerical calculations as described in section
\ref{sec:numcalc}. If only single vibron losses are present in the
system ($\gamma_{+}=0$, $\gamma_{2\pm}=0$), one finds from the
equation of motion (\ref{eq:nonlinQMEPsi}) for $\RDM$ that any
initial state $\rho(0)$ will decay to the ground state, i.e.,
$\rho(\infty)=\vert 0\rangle \langle 0 \vert$. The associated position
variance is $\mean{ \Delta q^2 }/q_0^2=1/2$.

On the other hand, if only nonlinear losses are present
($\gamma_{2+}=0$, $\gamma_{\pm}=0$), a general initial state
will relax to the steady state \cite{silo78,gikn93}
\be\label{eq:qubit_state}
\rho(\infty)=
\rho_{0,0}\vert 0 \rangle \langle 0 \vert + \rho_{0,1}\vert 0 \rangle
\langle 1 \vert+ \rho_{1,0}\vert 1 \rangle \langle 0 \vert+
\rho_{1,1}\vert 1 \rangle \langle 1 \vert, 
\ee
where the matrix elements in (\ref{eq:qubit_state}) depend on the
initial density matrix $\RDM(0)$. Since for
two-vibron losses parity is conserved,
the only nonzero matrix elements on the main
diagonal of $\rho(\infty)$ are
\begin{eqnarray}
\rho_{0,0}(\infty)&=&\psi_0(0,\infty)=\sum_{n\: {\rm even}}\rho_{n,n}(0)
\equiv P_{\rm even},\\ 
\rho_{1,1}(\infty)&=&\psi_1(0,\infty)=\sum_{n\:
  {\rm odd}}\rho_{n,n}(0) \equiv P_{\rm odd}\;. \nonumber
\end{eqnarray}
The steady state off-diagonal elements $\rho_{0,1}$
and $\rho_{1,0}$ are in general non-zero and can also be found from a
conservation law. From \eref{eq:nonlinQMEPsi} one gets
\bdm
\rho_{0,1}(\infty)=\rho_{1,0}^{*}(\infty)=\sum_{n=0}^{\infty}C_{2n}\psi_{2n}(1,0)
e^{\imath(\Omega+2\mu) t}, \edm
where the coefficients $C_{2n}$ are given in terms of Gamma functions
(see \ref{sec:app_zeroT_NLD})
\begin{equation}\label{eq:C2n}
  C_{2n} = \frac{\Gamma(\frac{1}{2} + n) 
    \Gamma(1-\imath \frac{\mu}{\gamma_{2-}})}{\sqrt{\pi}
    \Gamma(1 + n-\imath \frac{\mu}{\gamma_{2-}}) }\;.
\end{equation}
This expression is a generalization of the result of Simaan and Loudon
\cite{silo78} to the case of an anharmonic oscillator
($\mu\neq0$). Note, that unless $\mu=0$, the off-diagonal element
$\rho_{0,1}$ is time-dependent in the long time limit. Specifically,
for a system which is initially in a coherent state
$\ket{\psi(t=0)}=\ket{\alpha}$, i.e., with matrix elements
\begin{equation}
  \RDM_{n,m} (0) = \exp(-|\alpha|^2) \alpha^n \alpha^{*m}/\sqrt{n!m!}\;,
\end{equation}
the constants of motion are
\begin{eqnarray*}
  P_{\rm even} =  \exp(-|\alpha|^2) \cosh(|\alpha|^2)\;,\\
  P_{\rm odd} =  \exp(-|\alpha|^2) \sinh(|\alpha|^2)\;,
\end{eqnarray*}
and
\begin{equation}\label{eq:psi_t_even}
  \psi_{\rm even}(t)\equiv\rho_{0,1}(\infty) = \alpha^* \exp(-|\alpha|^2 -\imath \mu t) 
        {}_0F_1(1-\imath\frac{\mu}{\gamma_{2-}},|\alpha|^4/4)\;.
\end{equation}
Here, ${}_0F_1$ is the confluent hypergeometric limit
function \cite{we03}. For $\mu=0$ the latter becomes a Bessel function of
the first kind and the
expression given in \cite{silo78} is recovered.

\begin{figure}[t!]
  \centering
 \includegraphics[width=0.49\textwidth]
                      {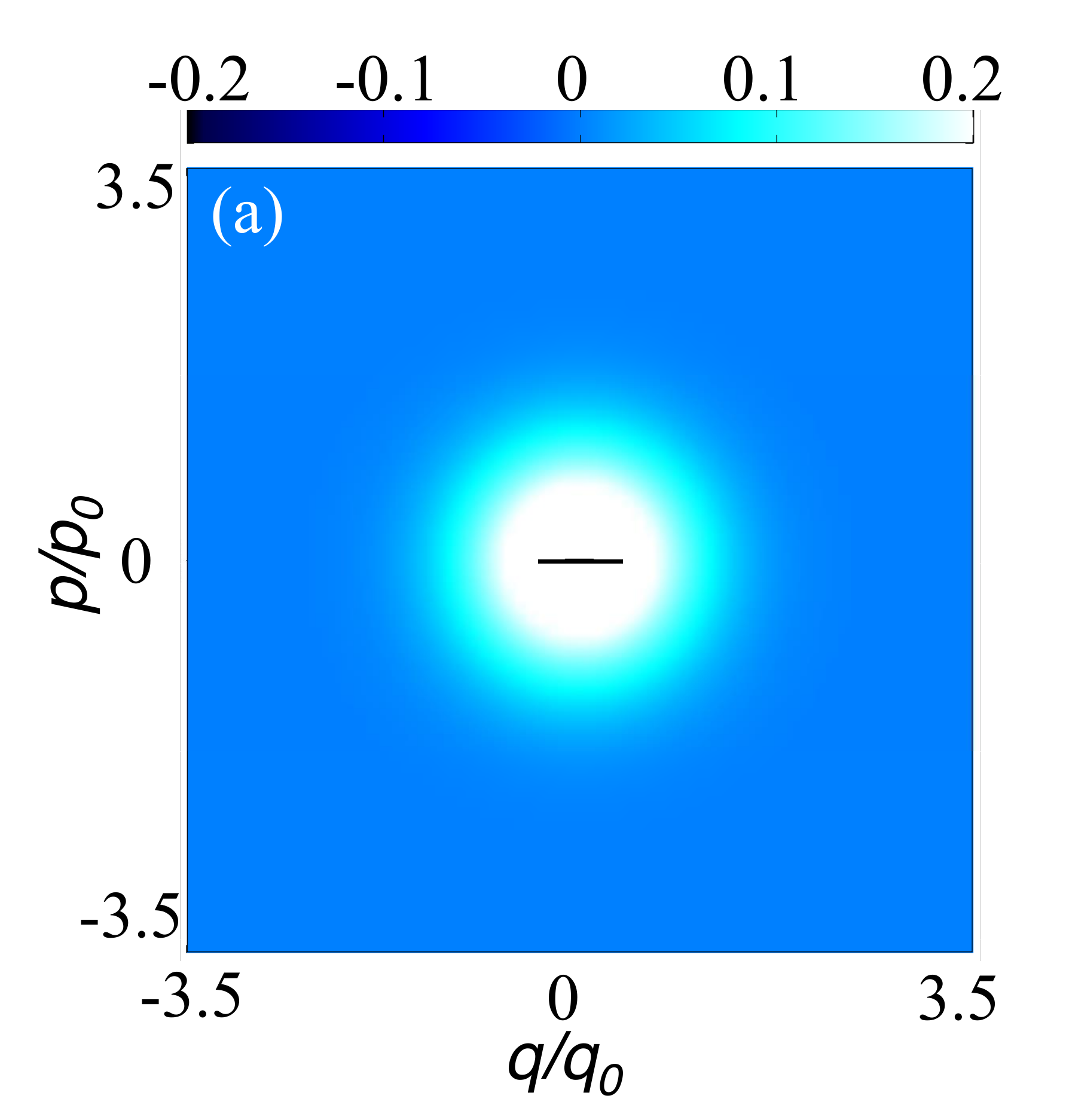}
      \includegraphics[width=0.49\textwidth]
                      {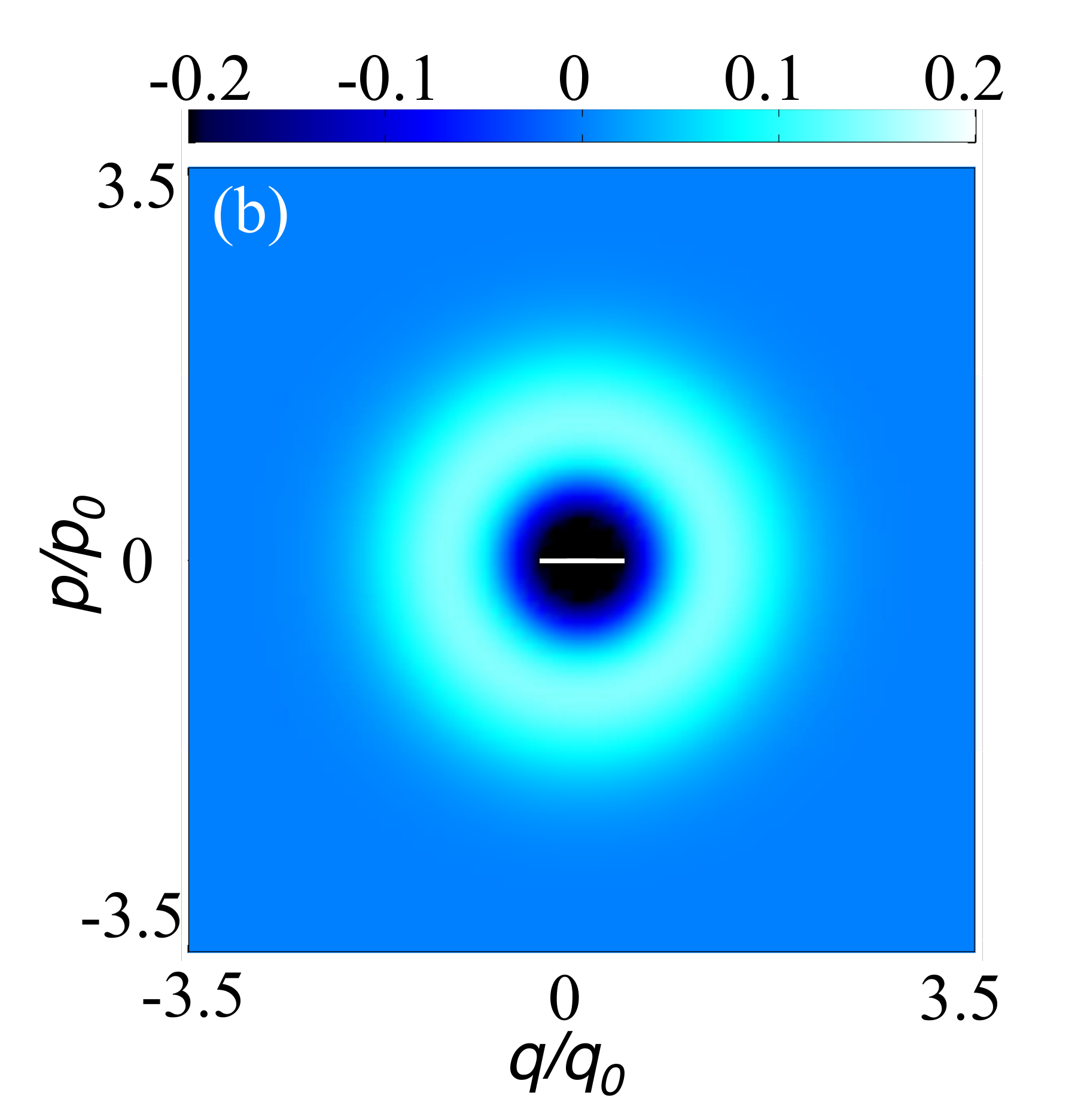}\\
      \includegraphics[width=0.49\textwidth]
                      {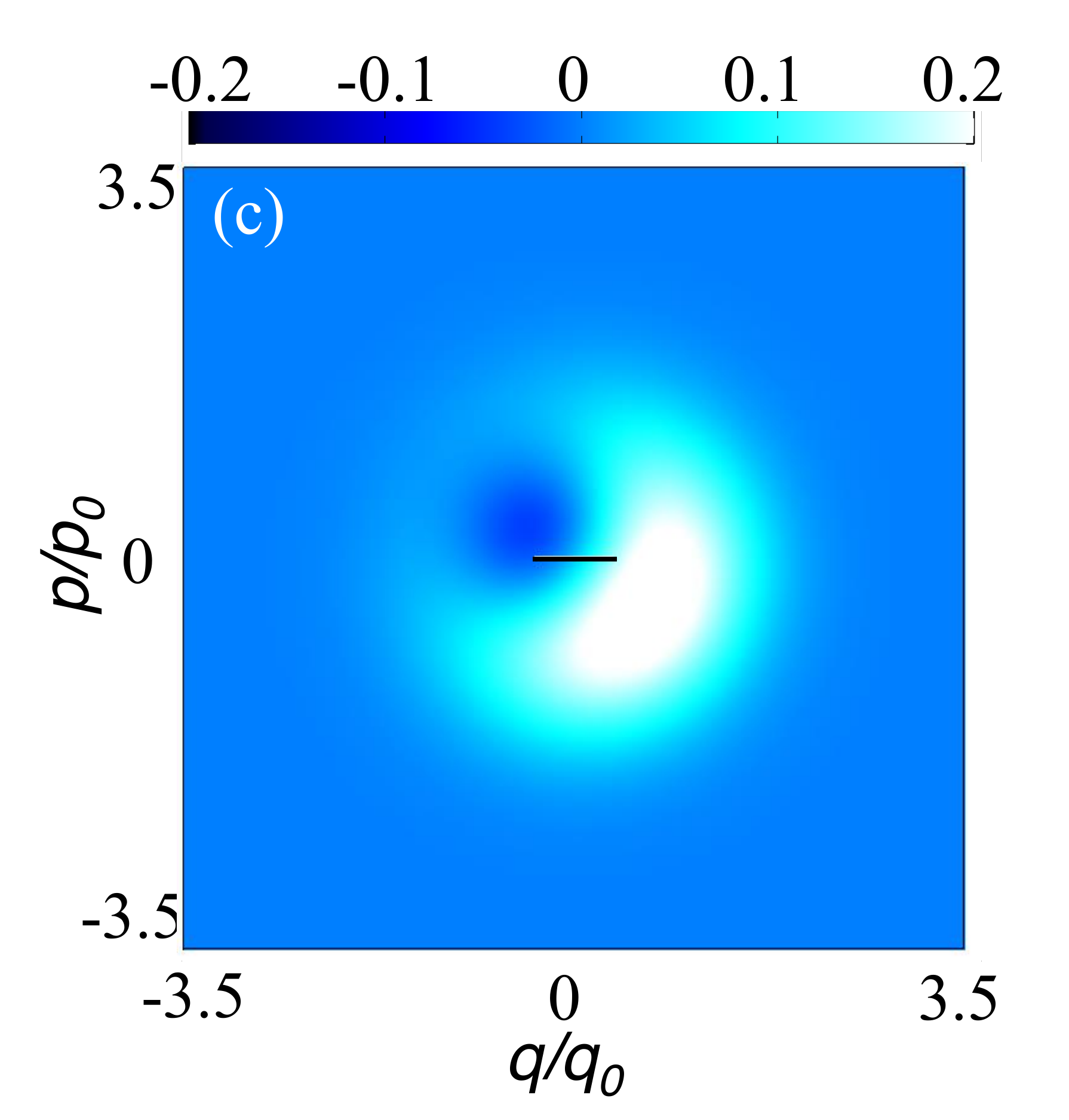}
      \includegraphics[width=0.49\textwidth]
                      {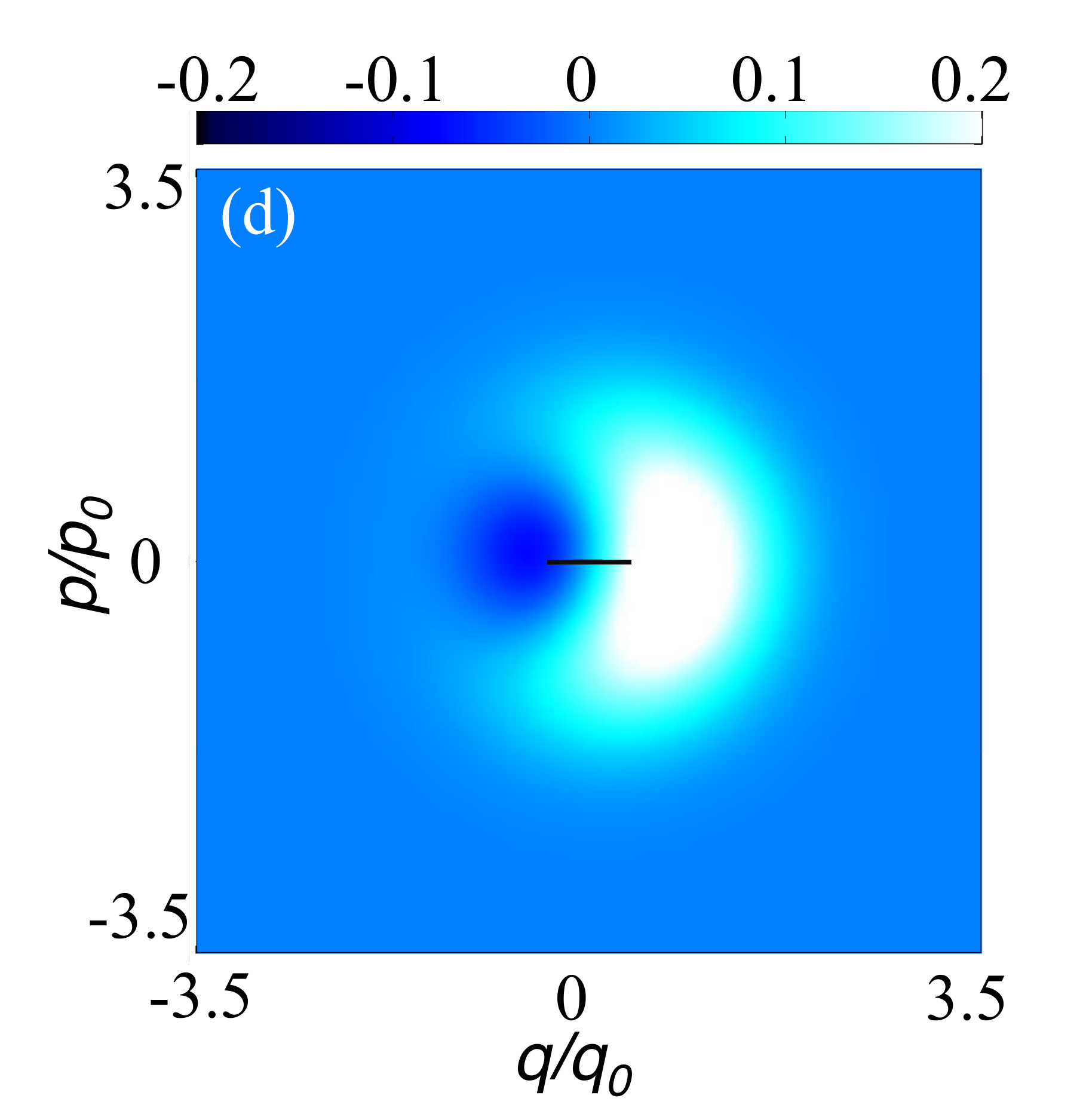}
\caption{Wigner distributions (\ref{eq:def_wigner}) of the two
  lowest number states and typical steady states resulting from
  nonlinearly damped, initial coherent states with $\alpha=1$ at $T=0$.
 (a)
  Ground state $\vert 0 \rangle \langle 0
  \vert$. The black line indicates the size of the state given by
  $\sqrt{\mean{\Delta q^2}}/q_0=1/\sqrt{2}$.  
(b) First excited number state
       $\vert 1 \rangle \langle 1 \vert$. The white line indicates the
       size of the ground state.  
(c) Weak NLD ($\gamma_{2-}/\mu=1/10$). 
The negative domains in the Wigner
distribution reflect the presence of coherences (non-zero
off-diagonals) in the steady state. For comparison the black line
shows the size of the ground state.  
(d) Strong NLD ($\gamma_{2-}/\mu=10$).
         The black line is included for the same purpose as in (c).
 \label{fig:wigner}}
\end{figure}
To illustrate the differences between the numerically calculated
states obtained by LD and NLD, their Wigner distributions
\cite{sc01}
\be\label{eq:def_wigner}
	W(q,p)=(2\pi)^{-1}\int d\xi~ e^{-\imath p \xi} 
	\langle q +\xi/2\vert \RDM(\infty) \vert q-\xi/2\rangle,
\ee
are displayed in figure \ref{fig:wigner}. For comparison also the
Wigner distributions of the ground and the first excited number state
are included in figures \ref{fig:wigner} (a) and (b). While the ground
state Wigner distribution is a Gaussian minimum uncertainty state, the
first excited number state is a quantum state with negative domains in
the Wigner distribution. The $\rho(\infty)$ resulting from NLD is a
non-classical state with both negative and positive values in the
Wigner distribution. This state is a mixed state which possesses
coherences in terms of nonzero off-diagonal elements $\rho_{0,1}$ in
the number basis (cf. \eref{eq:qubit_state}). 
In general it can be shown that the purity of this state is
  $\rm{Tr}[\rho^2]=1-2P_{even}P_{odd}+2\vert\rho_{0,1} \vert^2 <1$.
Figures \ref{fig:wigner}
(c) and (d) display typical snapshots of this steady state
for two different values of the ratio between the nonlinear damping
rate and the Kerr constant. By comparing the sizes of the stationary
states obtained by LD and NLD, it is seen that the latter occupy a
larger phase-space area. This suggests, that the position variance
$\langle \Delta q^2 \rangle$ is suitable for quantifying the
differences between the states resulting from these damping
mechanisms.

\begin{figure}[t!]
  \centering
      \includegraphics[width=0.49\textwidth]
                      {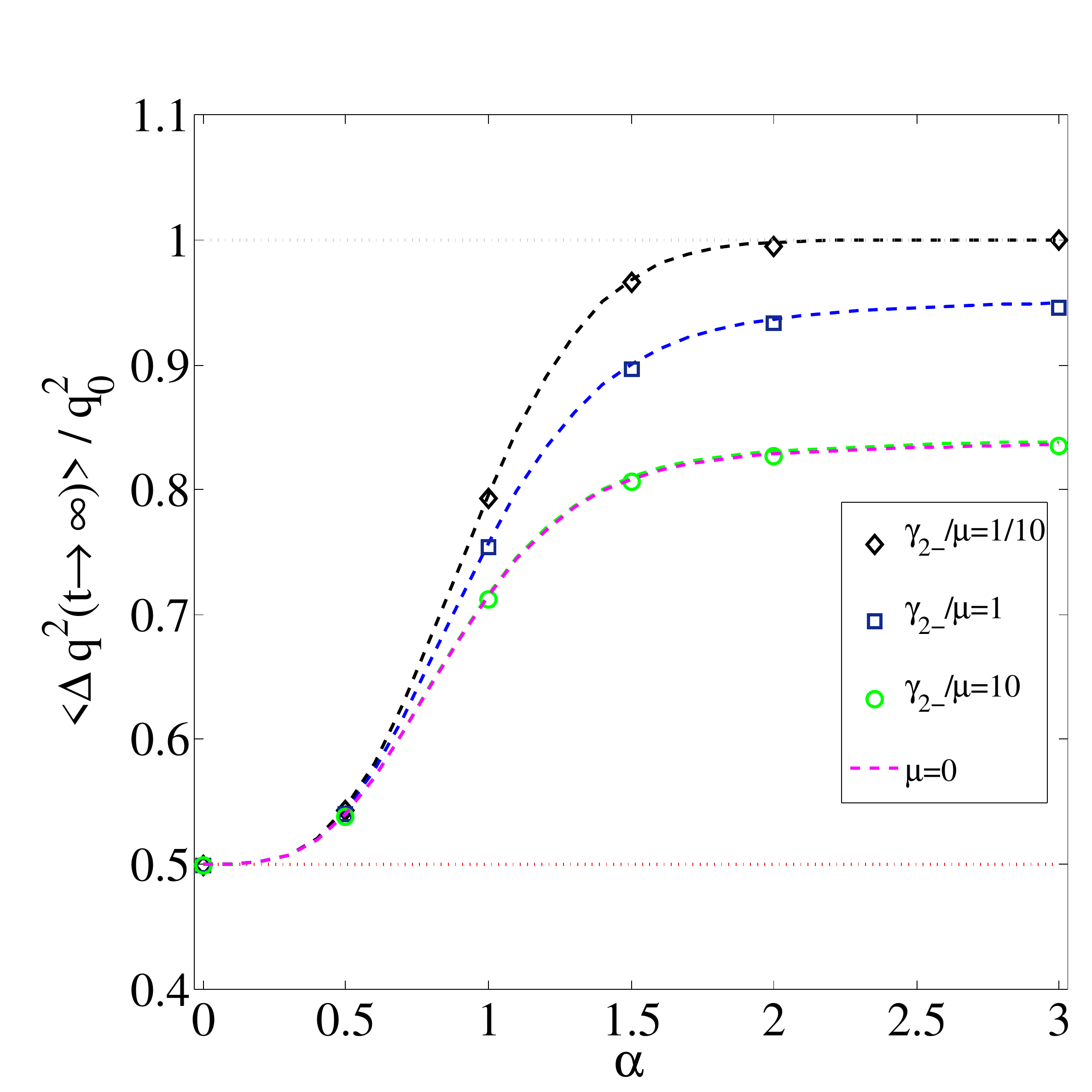}     
\caption{Zero temperature position variance of the steady state
  resulting from NLD as a function of the displacement amplitude
  $\alpha$ of the initial coherent state. The variance is averaged
  over one period $2\pi/\mu$. Dashed lines show the behavior according
  to \eref{eq:varT0analytic} for different values of the ratio
  $\gamma_{2-}/\mu$. Symbols denote the corresponding numerical
  results. The lower dotted line (red) indicates the variance of the
  ground state, resulting from LD. The upper dotted line (gray)
  indicates the upper variance limit, resulting from weak NLD for
  $\alpha\rightarrow \infty$.
  \label{fig:varT0NLD}}
\end{figure}
Using the definition \eref{eq:quad_def_rwa} and the results above, one
can evaluate the variance to be
\be\label{eq:varT0analytic}
 \langle \Delta q^2(\alpha)\rangle_{T=0}/q_0^2
 =\frac{1}{2} + e^{-|\alpha|^2}
  \sinh(|\alpha|^2) - \vert e^{-|\alpha|^2} \alpha^*
        {}_0F_1(1-\imath \mu/\gamma_{2-},|\alpha|^4/4)\vert^2 .
\ee
In figure \ref{fig:varT0NLD}, the variance according to
\eref{eq:varT0analytic} is shown as function of the displacement
amplitude $\alpha$ of initial coherent states for different values of
$\gamma_{2-}/\mu$ (dashed lines). The numerical results (symbols) are
in accordance with \eref{eq:varT0analytic}. For comparison, also the
variance of the ground state is included (red line), indicating its
independence on the initial displacement amplitude. For small initial
displacements, ($\alpha\lesssim 1$), the behavior of the variance is
governed by the coherences $\rho_{0,1}=\rho_{1,0}^*$, since
$\rho_{1,1}\to0$. For large amplitudes, ($\alpha\gg 1$), and weak NLD,
($\gamma_{2-}\ll \mu $), the variance saturates at $1$ (gray dotted
line). This is due to the ground and the excited states being equally
populated, $\rho_{0,0}=\rho_{1,1} = 1/2$, and the vanishing
off-diagonal elements, since the hypergeometric function converges to
zero. In the limit of strong NLD, ($ \gamma_{2-}\gg\mu$), and large
amplitudes, ($\alpha\gg1$), the hypergeometric function converges to a
non-zero value and the variance becomes $\mean{ \Delta q^2 } \to 1-
1/2\pi$, which is larger than the variance of the ground state.
 
\subsection{Finite temperature}\label{sec:res_finT_NLD}\label{sec:finT_stationary}
\begin{figure}[t!]
  \centering \includegraphics[width=0.53\textwidth]
             {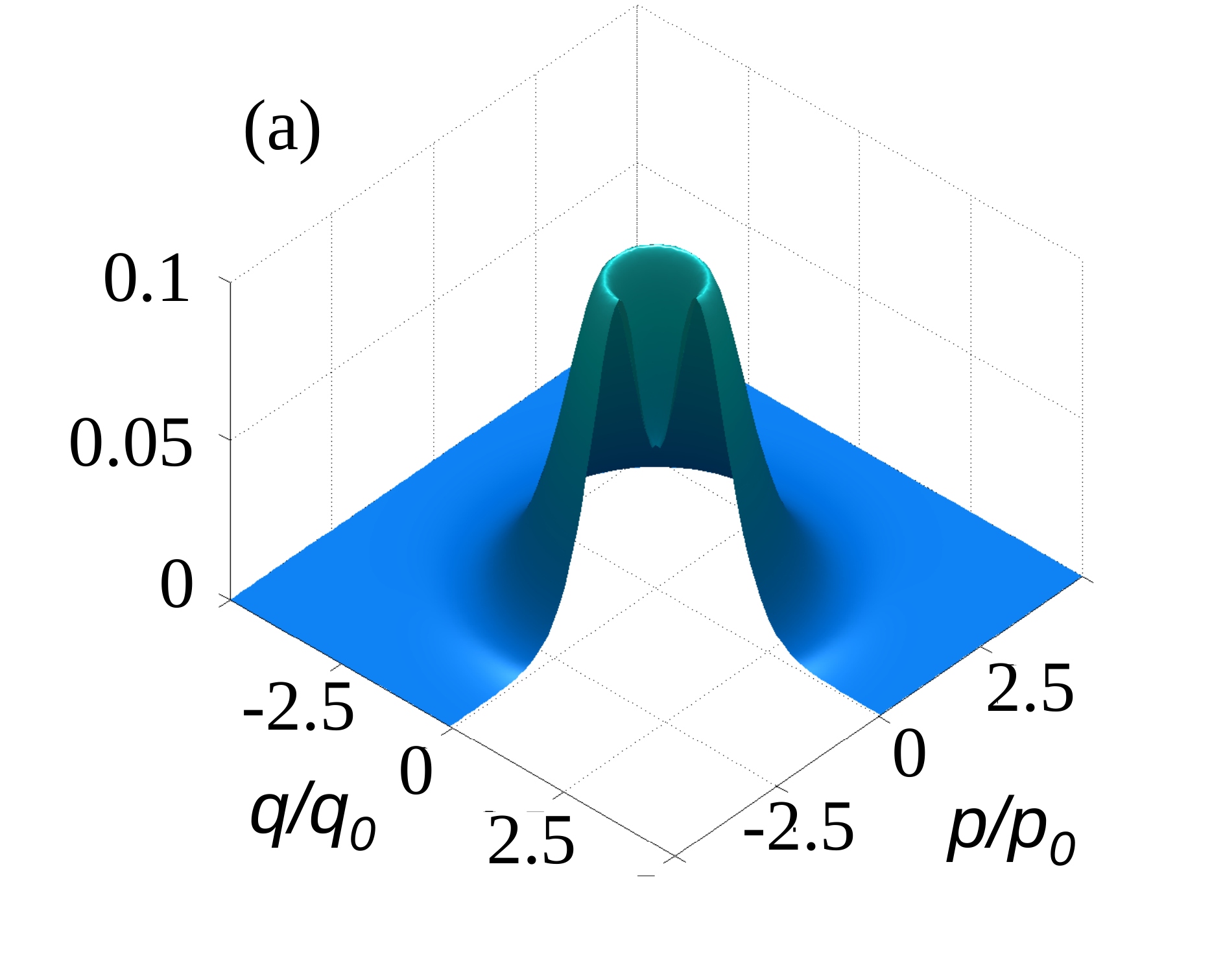}
             \includegraphics[width=0.46\textwidth]
                             {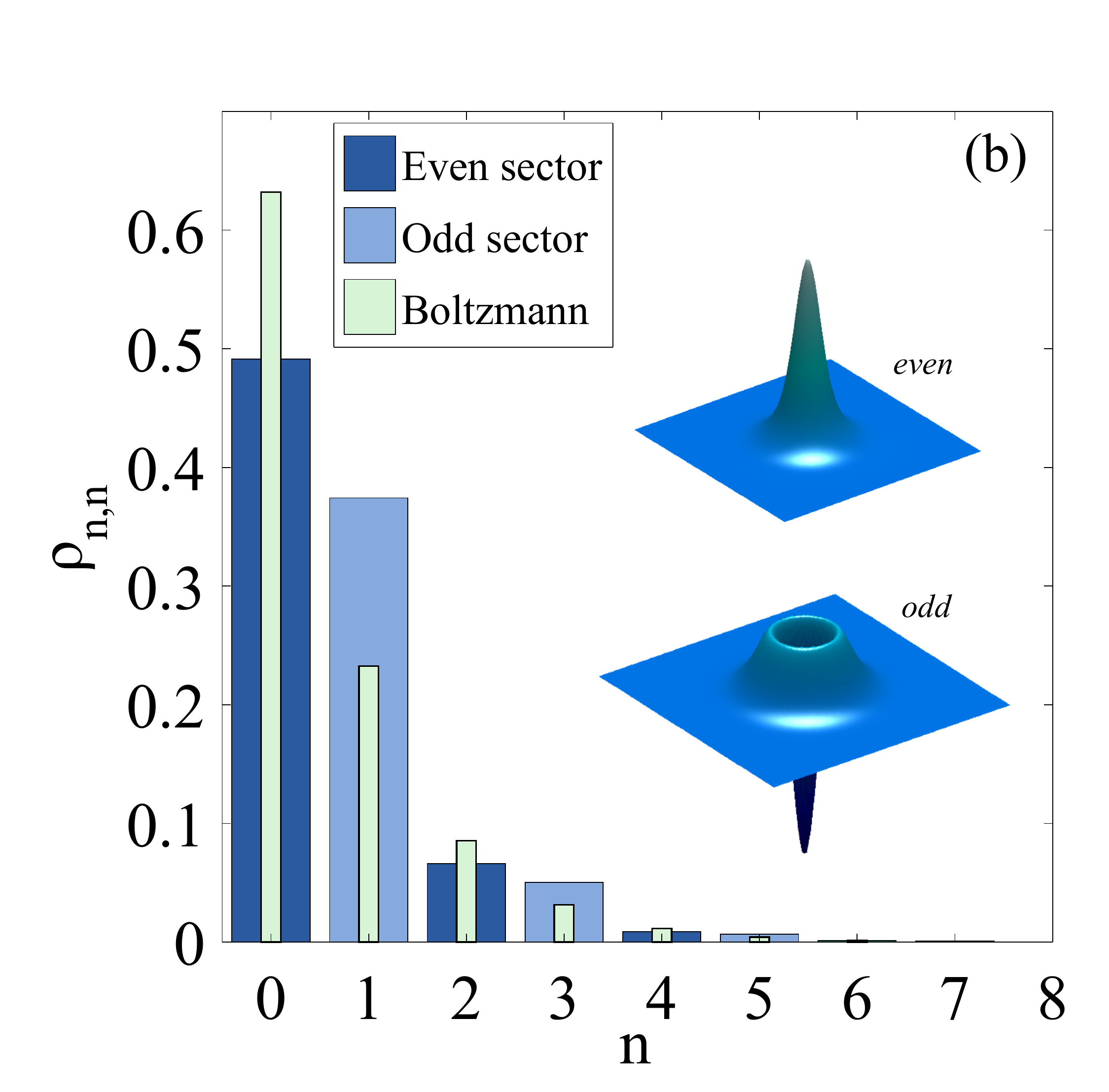}
\caption{(a) Wigner distribution \eref{eq:def_wigner} of the
  stationary state resulting from NLD at finite temperature with
  parameters $\gamma_{2-}=\mu,~\alpha=1,~k_{\rm B}T/\Omega=1$. A
  sector of the Wigner distribution is cut out for a better
  visualisation of the dip in the middle, which indicates the
  difference to a regular thermal state.
(b) Probability distribution of the even (dark blue bars) and odd
  (light blue bars) number state sectors of the stationary state shown
  in (a). The narrow bars represent the Boltzmann distribution of a
  thermal state described by \eref{eq:finT_LD}. The insets show the
  Wigner distributions of the even and odd sectors, respectively.
\label{fig:stat_state_finT}}
\end{figure}
Also at finite temperature $\rho(\infty)$ obtained by NLD will differ
from the state obtained by LD. For LD only, $\rho(\infty)$ is fully
characterized by the values of the diagonal matrix elements of the
density matrix,
\be\label{eq:finT_LD}
\rho_{n,n} = Z(\Omega)^{-1} e^{- n \Omega/(k_{\rm B} T) },
\ee
where $Z(\Omega)=\sum_n \exp[- n \Omega/(k_{\rm B} T)]$. Here, $k_{\rm
  B}$ is the Boltzmann constant and $T$ is the temperature of the
bath. This result is found from \eref{eq:nonlinQMEPsi} by taking
$\lambda=0$ and setting the time-derivative to zero. One obtains the
detailed balance condition, which leads to \eref{eq:finT_LD}. As we
have pointed out in section \ref{sec:RWA}, the RWA equations
\eref{eq:gen_lindblad} are valid for weak nonlinearities since the
bath is only probed at the frequency $\Omega$. Consequently, at finite
temperatures the stationary probability distribution calculated within
the RWA will not depend on the Kerr constant. However, for
sufficiently small temperatures, i.e., $\mu \mean{a^\dagger a} \ll
\Omega + \mu $, the approximation is valid as we will also show in the
following.

Since for nonlinear damping the two-vibron exchange conserves parity,
one finds that detailed balance holds for each parity sector
individually in this case. It can be shown (\ref{sec:app_fin_temp}) that
\begin{eqnarray} \label{eq:finT_NLD}
\rho_{2n,\: 2n} &=& Z(2\Omega)^{-1} e^{- 2n \Omega/(k_{\rm B} T)
        } P_{\rm even} \;,\\ 
\rho_{2n+1,\: 2n+1} &=& Z(2\Omega)^{-1} e^{-
          2n \Omega/(k_{\rm B} T) } P_{\rm odd}\;. \nonumber
\end{eqnarray}
Note that the stationary probability distribution at finite
temperature also depends on the initial state. This implies that
expectation values will in general also depend on $P_{\rm even}$ and
$P_{\rm odd}$. For example, the average number of vibrons in the
oscillator mode is given by
\begin{equation}\label{eq:finT_n}
  \mean{a^\dagger a} = \sum^\infty_{n=0} n \RDM_{n,n} = 2 n_{\rm
    B}(2\Omega) + P_{\rm odd}\;,
\end{equation}
which is derived in \ref{sec:app_fin_temp}.

Typical features of a stationary state obtained from nonlinear damping
at finite temperature are displayed in figure
\ref{fig:stat_state_finT}. Figure \ref{fig:stat_state_finT} (a) shows
the simulated Wigner distribution (see (\ref{eq:def_wigner})), with
parameters $\gamma_{2-}=\mu,~\alpha=1,~k_{\rm B}T/\Omega=1$. A sector
is cut out for better visualisation of its features. As discussed in
section \ref{sec:finT_stationary}, the thermal probability
distribution of a nonlinearly damped state is split into an odd and an
even parity sector. Each of the sectors has a Boltzmann-like
distribution (see \eref{eq:finT_NLD}). The shape of the Wigner
distribution reflects this parity dependence by having a dip centered
at the origin. It can be shown that
$W(0,0)=(P_{\rm even}-P_{\rm odd})/2\pi$. Therefore, the dip in the Wigner
distribution only depends on the initial state and not on
temperature. Hence, in the case of an initial coherent state $\vert
\alpha\rangle$, the dip will only depend on the initial oscillator
displacement. For large displacement amplitudes the difference between
the sectors becomes smaller, and $W(0,0)\rightarrow 0$. If the initial
state had an odd parity, the dip of the Wigner distribution would
attain a negative value.

Figure \ref{fig:stat_state_finT} (b) shows the Boltzmann-like
distributions of the odd and the even sectors of the stationary state
discussed above. The narrow bars represent the distribution of the
thermal state obtained by linear damping (see \eref{eq:finT_LD}). The
insets show the Wigner distributions of the even and odd sectors.

\begin{figure}[t]
  \centering
      \includegraphics[width=0.49\textwidth]
                      {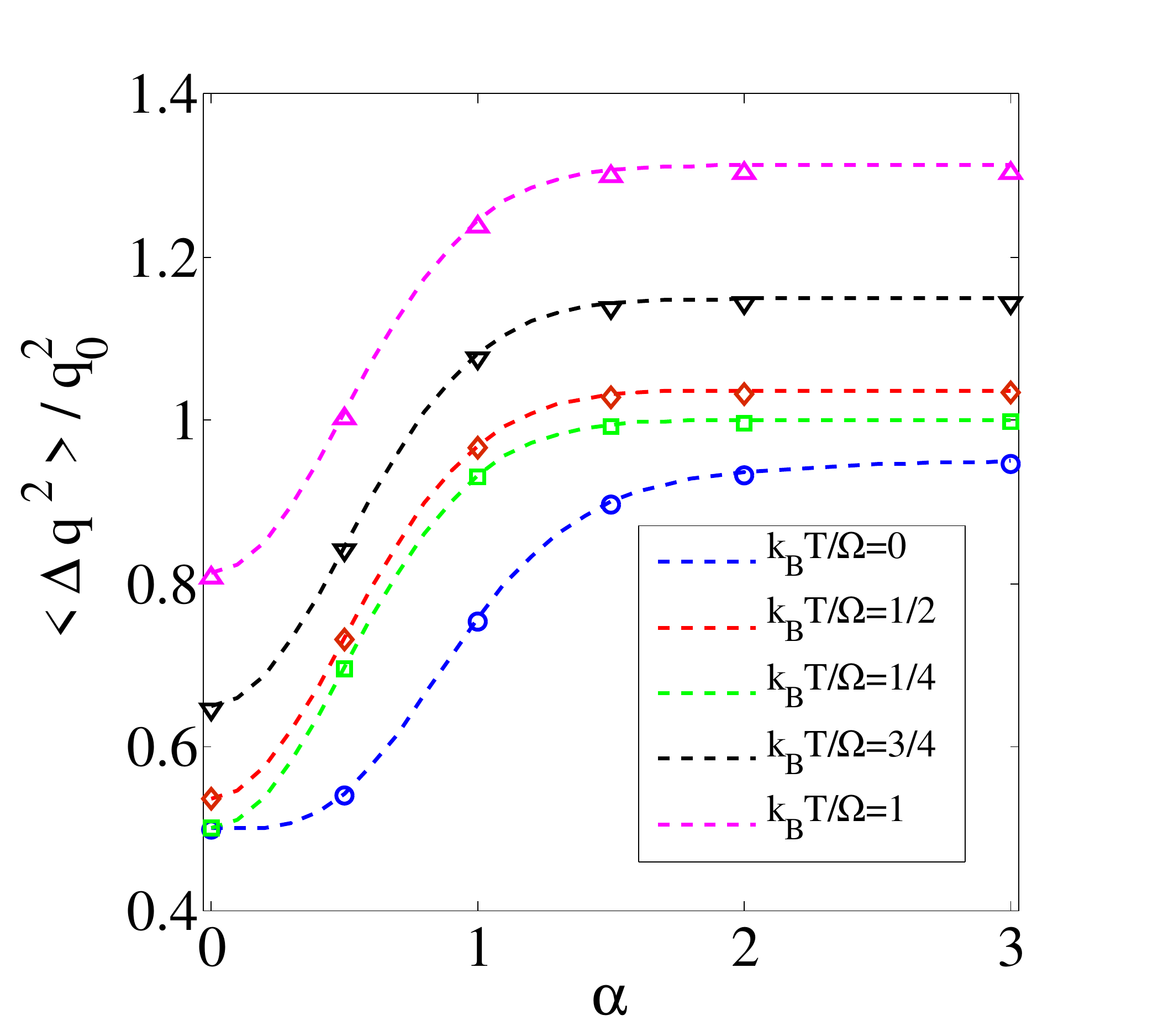}     
\caption{Finite temperature position variance of the stationary state
  as a function of the displacement amplitude $\alpha$ of the initial
  coherent state. The variance is averaged over one period
  $t=2\pi/\mu$. Dashed lines show the behavior according to
  \eref{eq:var_finT_NLD} for $\gamma_{2-}=\mu$ and different
  temperatures. Symbols denote the corresponding numerical
  results.\label{fig:finT_statvar}}
\end{figure}

As the Wigner distribution of the finite temperature stationary state
displays quantum signatures, so does its variance. Figure
\ref{fig:finT_statvar} shows the stationary state variance as function
of the displacement amplitude $\alpha$ of the initial coherent state
for several temperatures ($k_{\rm B}T/\Omega\in [0, 1/4, 1/2, 3/4, 1]$
and $\gamma_{2-}=\mu$). Dashed lines correspond to the analytical
expression given by (\ref{eq:var_finT_NLD}). Numerical calculations
(symbols) are shown for the same parameters. The variance is seen to
always be larger than the zero temperature variance, displayed in
figure \ref{fig:varT0NLD}. This is due to an additional contribution
from the Boltzmann distribution and the vanishing off-diagonal
elements of the density matrix (see section \ref{sec:finT_decay}). The
variance is still dependent on the initial displacement and saturates
at $\langle\Delta q^2 \rangle/q_0^2 =1+2n_{\rm B} > 1$ for large
$\alpha$, which is in contrast to $k_{\rm B}T/\Omega=0$ for which
$\langle \Delta q^2 \rangle/q_0^2 \leq 1$.

%
\section{Relaxation Dynamics}\label{sec:relaxation}

\subsection{Zero temperature, NLD}
\begin{figure}[t]
\begin{minipage}{0.7\textwidth}
  \centering
             \includegraphics[width=0.8\textwidth]
                             {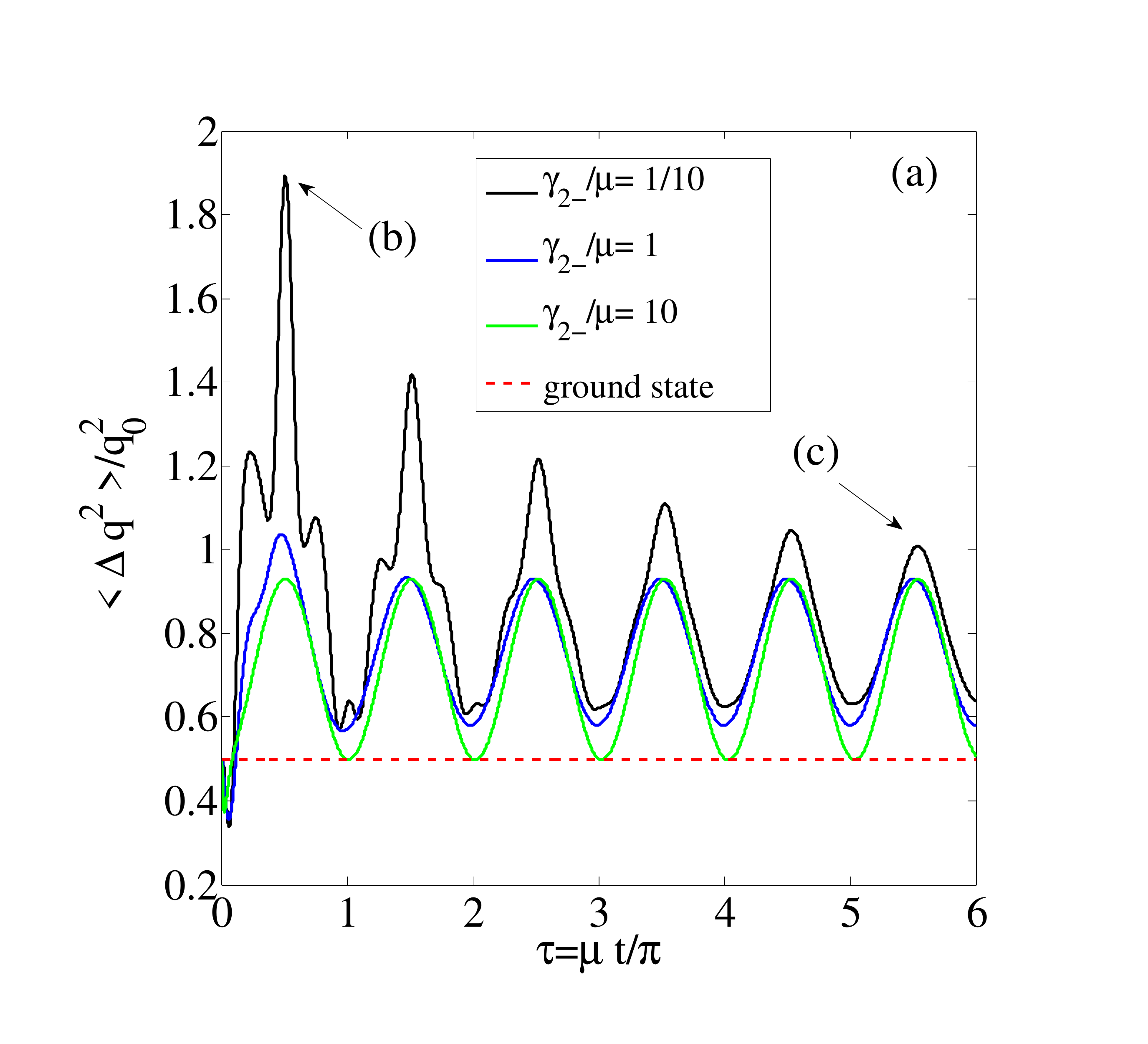}
\end{minipage}
\begin{minipage}{0.3\textwidth}
  \centering
  \vspace{0pt}
    \includegraphics[width=.8\textwidth]{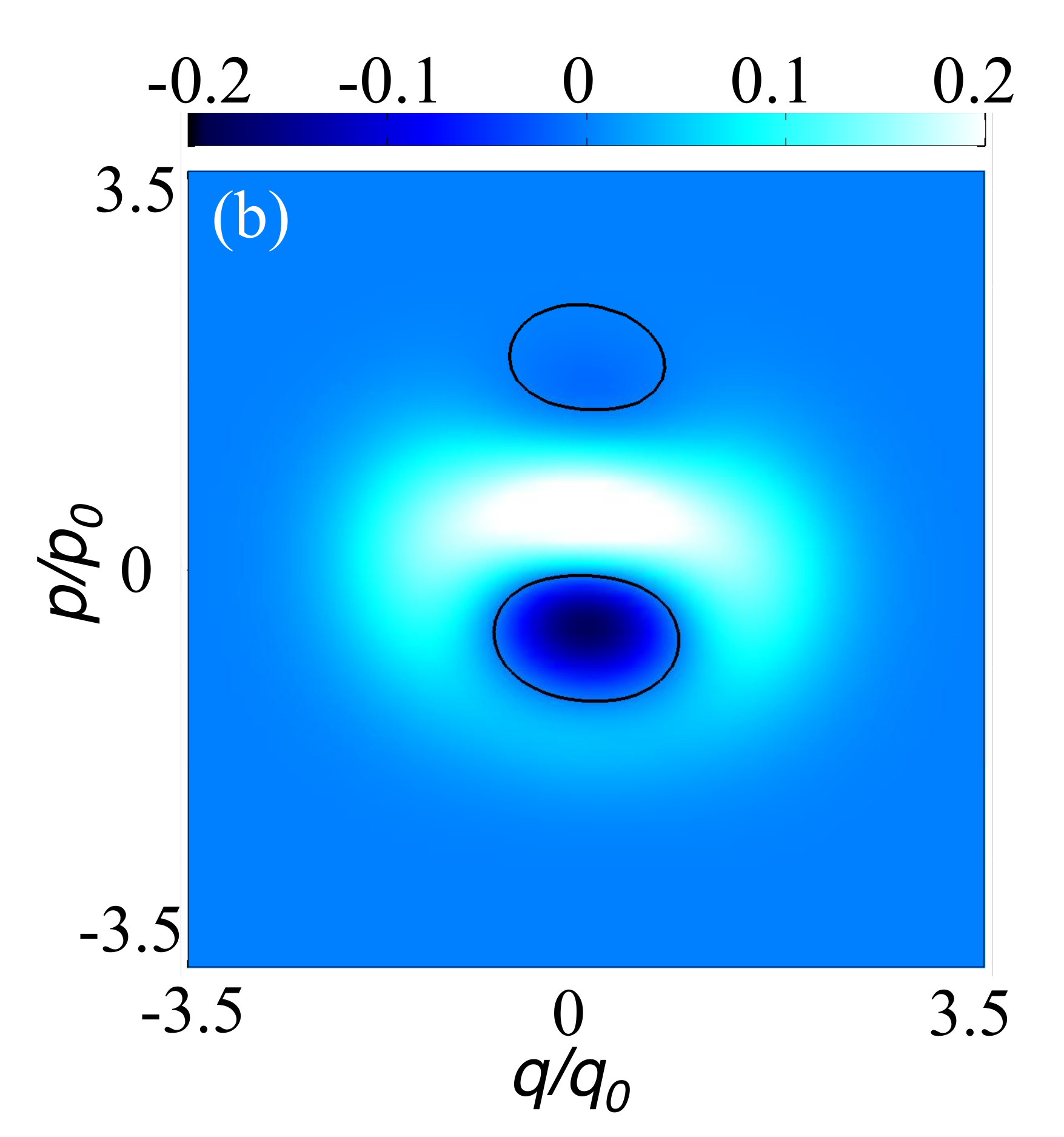}\hfill
    \includegraphics[width=.8\textwidth]{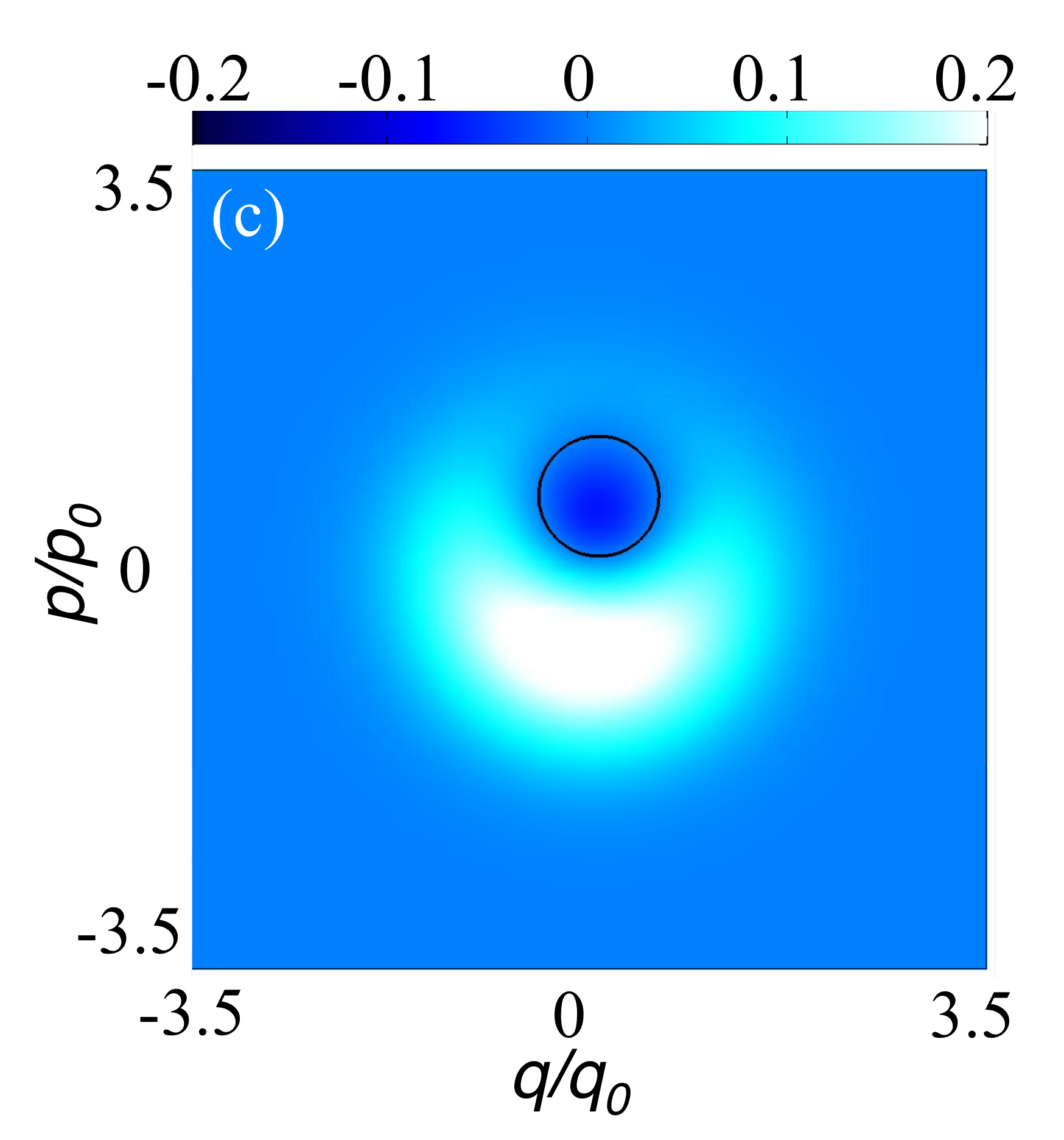}
  \end{minipage}
\caption{(a) Time evolution of the position variance of an initial
  coherent state with $\alpha=1$ at $T=0$ for different NLD
  strengths. The dashed line (red) indicates the variance of the
  ground state. (b) Wigner distribution $W(q,p)$ of the decohered
  Yurke-Stoler cat state attained at $\tau=1/2$. (c) Wigner
  distribution $W(q,p)$ of the state in (b) at a later time,
  $\tau=11/2$, reaching the steady state. Black contours
  encircle domains where $W<0$.\label{fig:var_relaxT0}}
\end{figure}

We now investigate the relaxation dynamics of initial coherent states
which evolve into the steady states discussed in the previous
section. Figure \ref{fig:var_relaxT0} (a) shows the numerically
calculated variance $\langle \Delta q^2 \rangle$ as function of
rescaled time $\tau=\mu t/\pi$ for several values of the ratio
$\gamma_{2-}/\mu$. The variance is sampled at multiples of the period
$\Omega/(2\pi)$, and therefore fast oscillations are not seen. For
weak NLD ($\gamma_{2-}/\mu \lesssim 10$), the short time dynamics
($\tau\lesssim 1$) is governed by the Kerr constant
$\mu$. This is reflected in the initial increase of variance. The
maximum variance, at which the Yurke-Stoler cat state is first
expected to appear \cite{yust86,voki+12}, is attained at
$\tau\approx 1/2$. The variance of this state is given by
$\langle \Delta q^2 \rangle_{\rm cat}=q_0^2(4\alpha^2+1)/2$, which is
always larger than the variance of a coherent state for a nonzero
$\alpha$. As seen in the figure \ref{fig:var_relaxT0} (a), the shape of
the variance peaks changes in time as the cat states decohere by
NLD. Figure \ref{fig:var_relaxT0} (b) shows the Wigner distribution
sampled at $\tau=1/2$, displaying a cat state under influence
of NLD. Some of its original features are still visible, like the
remnants of the negative interference domains.

After $\tau=1$, the system gradually relaxes to the state
\eref{eq:qubit_state}. In figure \ref{fig:var_relaxT0} (a) this is
reflected by steady oscillations around a mean value of $\langle
\Delta q^2 \rangle$ which is, as discussed in section
\ref{sec:statzeroT}, larger than $1/2$. The oscillations in $\langle
\Delta q^2 \rangle$ are caused by the complex valued, oscillating
coherences of $\rho_{0,1}$. The relaxation dynamics described above
becomes faster with increasing ratio $\gamma_{2-}/\mu$. Figure
\ref{fig:var_relaxT0} (c) shows the Wigner distribution of the state
sampled at $\tau=11/2$, where the resemblance to the steady
state in figure \ref{fig:wigner} (c) is clearly seen.

\subsection{Zero temperature, LD and NLD}\label{sec:res_zeroT_LD_NLD}
As shown in the previous section, NLD at $T=0$ leads to the formation
of the non-classical state \eref{eq:qubit_state}. In the additional
presence of LD the generation of this state will be affected and
eventually the ground state will be reached. In the following we
discuss the resulting decay of the variance, which can be used to
extract information about the nature of the damping mechanism and the
state initially created by NLD.
\begin{figure}[t]%
  \centering \includegraphics[width=0.49\textwidth]
             {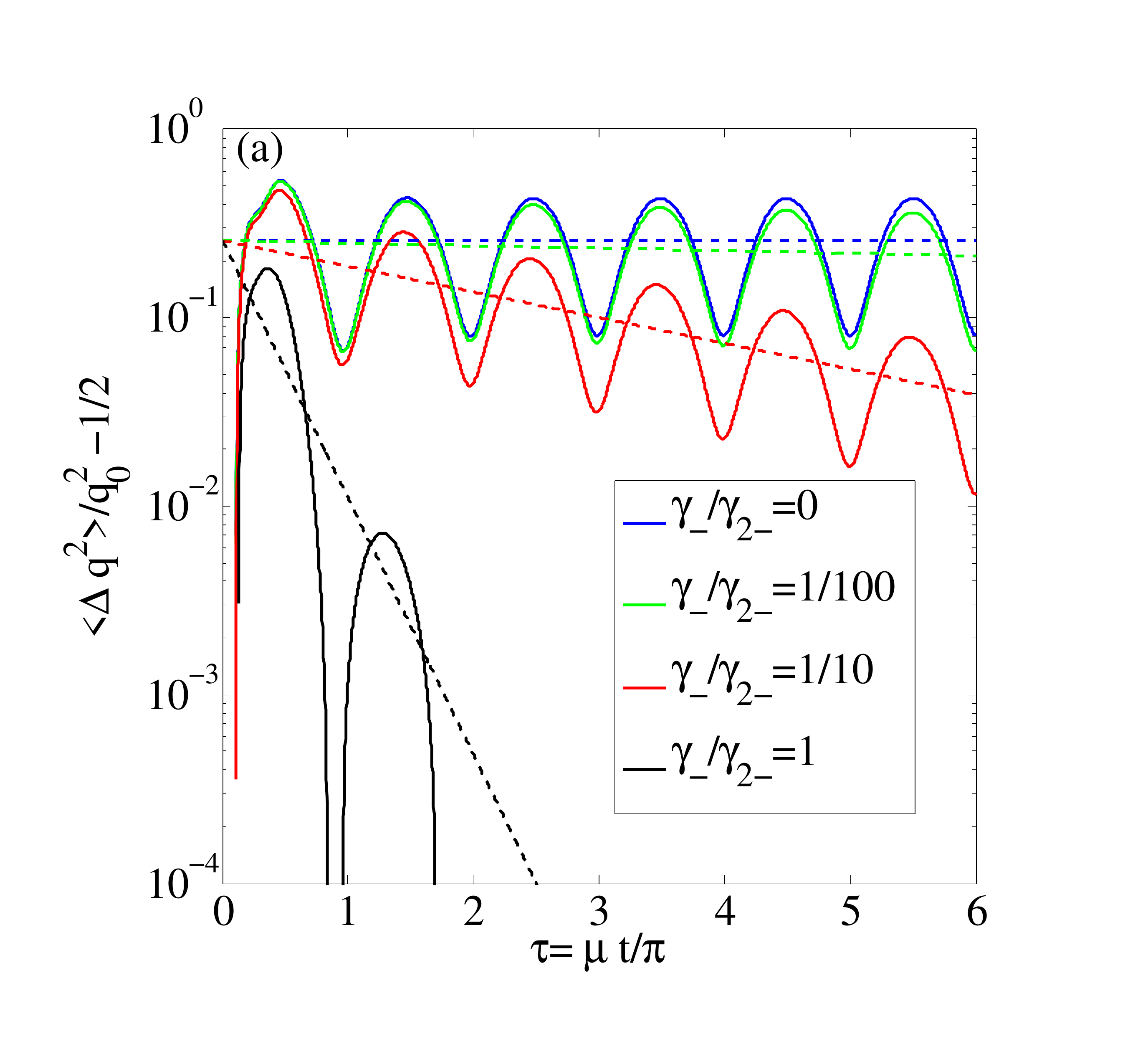}
  \centering \includegraphics[width=0.49\textwidth]
             {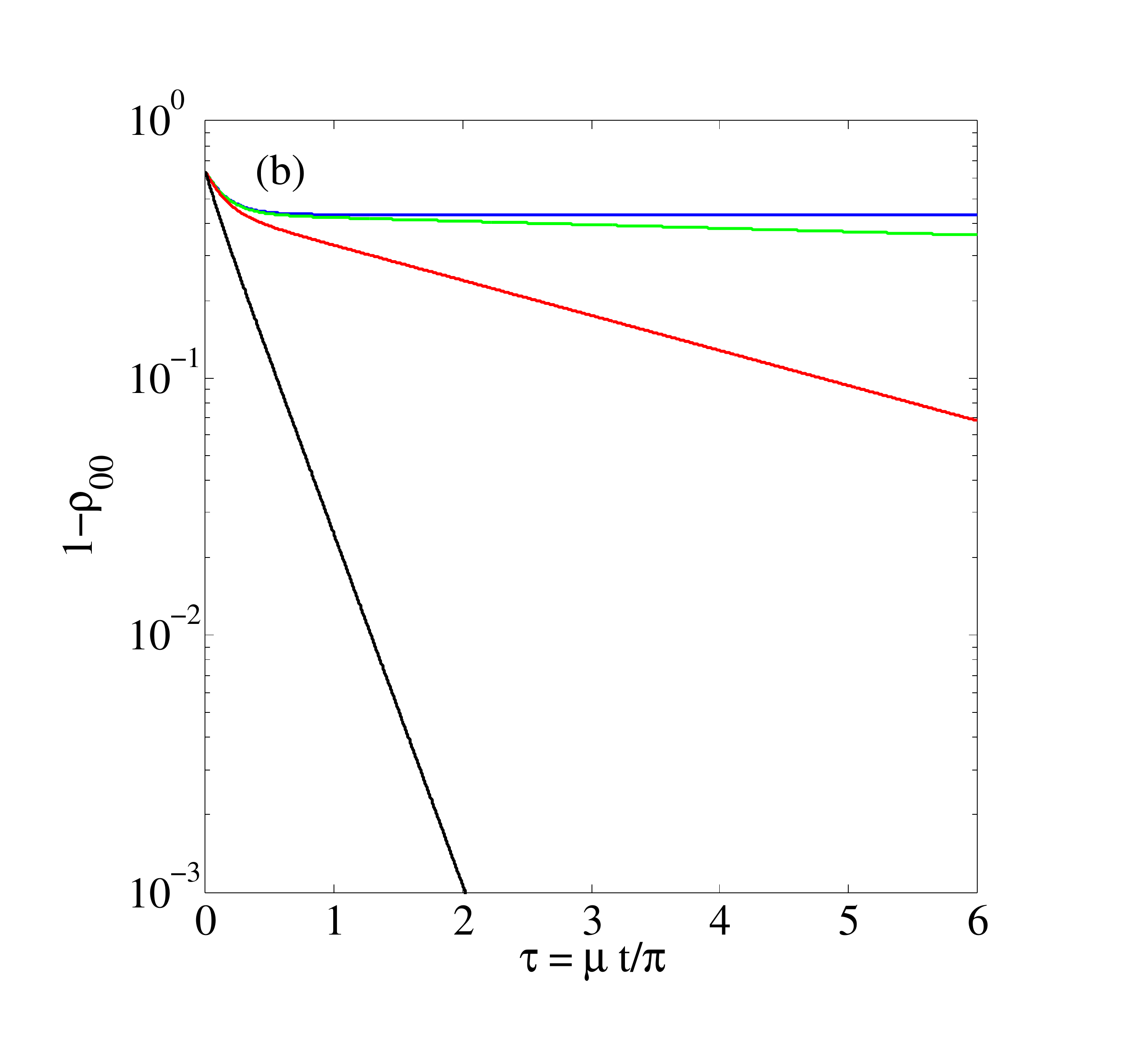}\\
  \centering \includegraphics[width=0.49\textwidth]
             {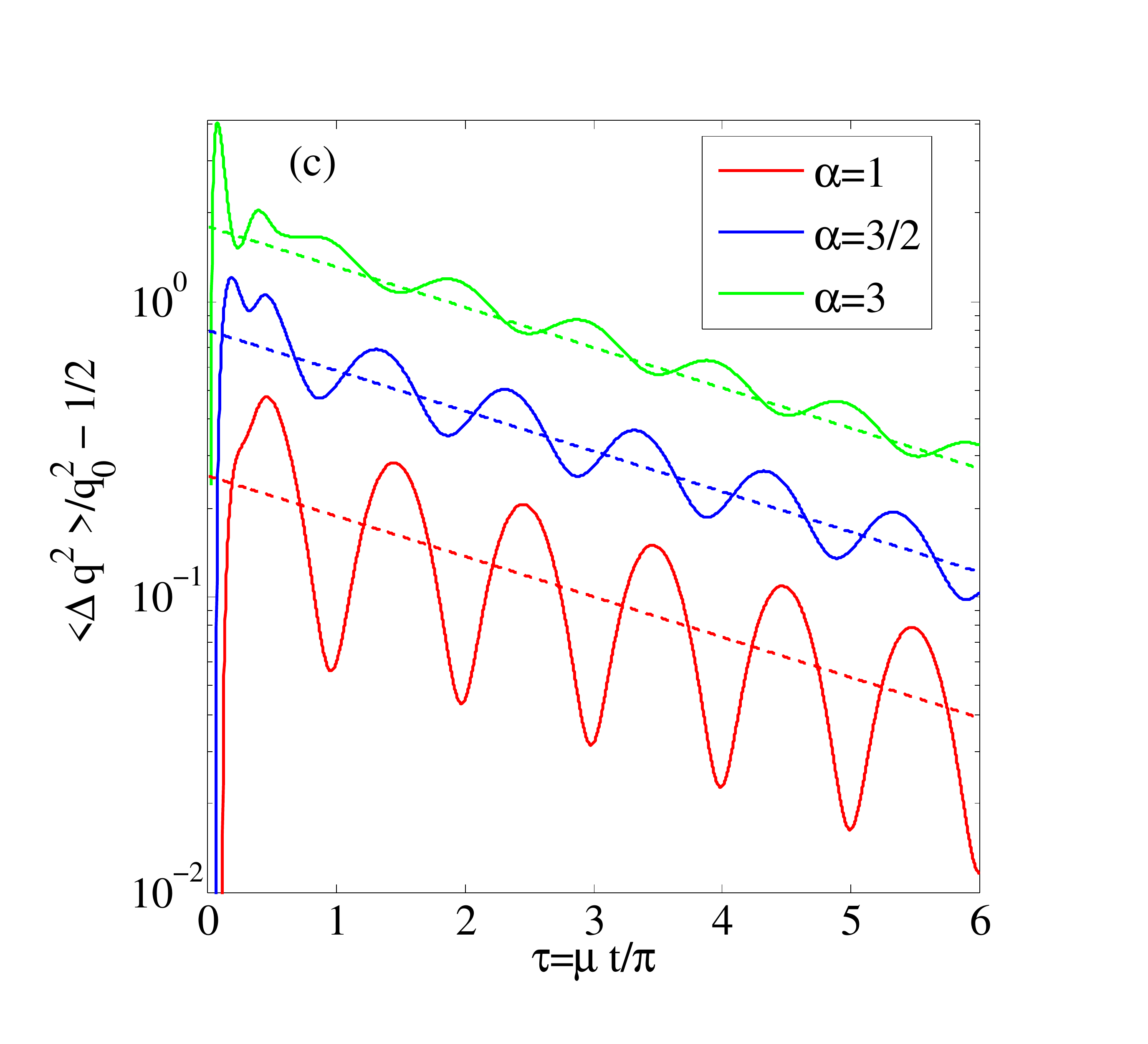}
\caption{ (a) Time-dependence of the position variance of a linearly
  and nonlinearly damped mechanical mode for $\alpha=1$ and
  $\gamma_{2-}=\mu$. Full lines denote different values of
  $\gamma_{-}/\gamma_{2-}$. Dashed lines show the function
  \eref{eq:var_zeroT_LD_NLD}. (b) Time-dependence of $(1-\rho_{0,0})$
  with parameters corresponding to those of the main figure.  (c) Time
  dependence of $\langle\Delta q^2(\tau) \rangle/q_0^2-1/2$ for
  $\gamma_{2-}=\mu$ and $\gamma_{-}/\gamma_{2-} =1/10$ for different
  initial displacement amplitudes. Dashed lines indicate the behavior
  according to \eref{eq:var_zeroT_LD_NLD}. For visual convenience, the
  graphs corresponding to $\alpha=3$ are multiplied by a factor of
  $2$.
  \label{fig:varT0NLD+LD_alpha1}}
\end{figure}
We introduce the ratio $\gamma_{-}/\gamma_{2-}$ as a measure of
the relative dissipation strength, and set
$\gamma_{2-}=\mu$. 

In presence of LD at $T=0$, \eref{eq:nonlinQMEPsi} leads to
(for $\lambda=0,1$)
\begin{eqnarray}\nonumber
\frac{\partial}{\partial t} \rho_{n,n}(t) &=& \gamma_{-}
(n+1)\rho_{n+1,n+1} + \gamma_{2-} (n+2)(n+1) \rho_{n+2,n+2}\\
	&& - \left[
  \gamma_{2-} n(n-1) + \gamma_{-} n \right]\rho_{n,n}\;,
\\ \nonumber
\frac{\partial}{\partial t} \psi_n(1,t) &=&
\gamma_{-} (n+1)\psi_{n+1} + \gamma_{2-} (n+2)(n+1)
\psi_{n+2} \\
&& -\left[ -\imath \mu (1+2n) + \gamma_{2-}
  n^2 + \gamma_{-} (n + \frac{1}{2}) \right] \psi_n\;.
\end{eqnarray}
It is seen that only $\rho_{0,0}$ is non-zero in the stationary limit,
i.e., $\rho_{1,1}$ and $\psi_n$ decay for sufficiently large times. In
the limit of strong NLD, $\gamma_-\ll\gamma_{2-}$, one finds that the
populations $\rho_{n,n}$ for $n>1$ rapidly decay due to two-vibron
losses, but $P_1$ is only decreased via one-vibron losses. Hence,
$\dot{\rho}_{1,1} \approx -\gamma_- \rho_{1,1}$ and $\rho_{1,1}(t)
\approx e^{-\gamma_- t}P_{\rm odd}$. Here it is assumed that the
initial decay on the time-scale $1/\gamma_{2-}$ is not influenced by
the LD and $\rho_{0,0}(0)\approx P_{\rm even}$ and
$\rho_{1,1}(0)\approx P_{\rm odd}$. Similarly, one finds that
$\psi_0(1,t) \approx \exp[(\imath\mu - \gamma_{-}/2)t]\psi_{\rm
  even}(0)$.  Therefore, the position variance will decay to the
minimum uncertainty value $1/2$ according to
\be\label{eq:var_zeroT_LD_NLD} \mean{\Delta q^2}/q_0^2 = \frac{1}{2} +
\left[ P_{\rm odd} - |\psi_{\rm even}(0)|^2 \right] e^{-\gamma_- t}\;.
\ee

Figure \ref{fig:varT0NLD+LD_alpha1} (a) shows the time evolution of the
numerically calculated $\langle \Delta q^2 \rangle$ for different
values of the ratio $\gamma_{-}/\gamma_{2-}$ and $\mu=\gamma_{2-}$.
For $\gamma_{-}/\gamma_{2-}\lesssim 1/10$, the evolution of the
variance for short times $\tau\lesssim 1$ resembles the result shown
in figure \ref{fig:var_relaxT0}. The short-time dynamics is again
influenced by the Kerr constant, while the long-time behavior, ($\tau
\gg 1$) is showing the relaxation of the state created by NLD. In
this stage the variance is exponentially decaying with the decay being
faster for larger values of $\gamma_{-}/\gamma_{2-}$.
In figure \ref{fig:varT0NLD+LD_alpha1} (b) the time evolution of
$(1-\rho_{0,0})$ is shown, confirming the decay to the ground
state. The parameters are the same as in \ref{fig:varT0NLD+LD_alpha1}
(a). For equally strong LD and NLD, ($\gamma_{-}=\gamma_{2-}$), an
exponential decay is seen for the entire time interval. For LD weaker
than NLD, ($\gamma_{-}<\gamma_{2-}$) and $\tau\lesssim 1/2$, the
decay is initially not exponential, whereas for $\tau\gtrsim 1$
the ground state population displays exponential increase.

Figure \ref{fig:varT0NLD+LD_alpha1} (a) suggests the concept of
sequential relaxation for weak LD. The first stage of the sequence ($0
< \tau \lesssim 1$), is mainly dominated by NLD and the Kerr
constant, which leads to the formation of the state in
\eref{eq:qubit_state}. The second stage ($\tau \gtrsim 1$), is
mainly influenced by the relaxation to the ground state because of
LD. In other words, the state in \eref{eq:qubit_state} is initially
created by NLD, and then decays due to LD. Accordingly, the average
variance in the second relaxation stage is expected to be given by
\eref{eq:var_zeroT_LD_NLD}, which is displayed in figure
\ref{fig:varT0NLD+LD_alpha1} (a) (dashed lines). It can be seen that
the idea of sequential relaxation is quantitatively correct for
$\gamma_{-}\ll\gamma_{2-}$. Furthermore, \eref{eq:var_zeroT_LD_NLD}
implies that the decay rate, which is equal to $\gamma_{-}$, is
independent of the initial amplitude $\alpha$. This is shown in figure
\ref{fig:varT0NLD+LD_alpha1} (c) where the time evolution of $\langle
\Delta q^2(\tau) \rangle$ is plotted for $\alpha=1,~3/2$ and $3$
($\gamma_{2-}=\mu$, $\gamma_{-}/\gamma_{2-}=1/10$).

\subsection{Finite temperature, NLD}\label{sec:finT_decay}
While at $T=0$ two-vibron losses facilitate the creation of the
non-classical state \eref{eq:qubit_state}, thermal two-vibron
excitations will contribute to its decoherence. From
\eref{eq:nonlinQMEPsi} one sees that a finite rate $\gamma_{2+}$
couples the states ($n=0,1$) to other excited
states. These thermal excitations will eventually destroy the
coherence of the non-classical states and the off-diagonal matrix
elements will vanish. To quantify this decay we consider low
temperatures, such that $\gamma_{2+}\ll\gamma_{2-}$. To obtain the
relaxation rate to lowest order in $\gamma_{2+}$ it is sufficient to
study excitations from $n=0$ to $n=2$. Consequently, one gets from
\eref{eq:nonlinQMEPsi} the coupled equations
\begin{eqnarray}\label{eq:psi_dot}
	\dot{\psi}_0(1,t) & = & (\imath \mu - 4\gamma_{2+})\psi_0
        +2\gamma_{2-}\psi_2, \\ \dot{\psi}_2(1,t) & = &
        6\gamma_{2+}\psi_0+( 5\imath\mu -16\gamma_{2+}-
        4\gamma_{2-})\psi_2\;,\nonumber
\end{eqnarray}
supplemented by initial conditions for $\psi_0(1,0)$ and
$\psi_2(1,0)$. For small $\gamma_{2+}$ we can assume that these
initial values correspond to the values of the state in
\eref{eq:qubit_state}, which is the steady state solution for
$\gamma_{2+}=0$. This amounts to setting
\begin{equation}
\psi_0(1,t=0)=\psi_{\rm even}(0),\quad
\psi_2(1,t=0)=0\;.
\end{equation}
The system \eref{eq:psi_dot} can be solved for $\psi_0(1,t)$,
\begin{eqnarray} \label{eq:psi0_decay_NLD}
\psi_0(1,t)&=&\Big[\cosh(2\chi
  t) + \Big(\frac{3\gamma_{2+}-\imath \mu +
    \gamma_{2-}}{\chi}\Big)\sinh(2\chi t)\Big] \\
  && \times \exp[-(2\gamma_{2-} +
  10 \gamma_{2+} - 3 \imath\mu)t]\psi_{\rm even}(0)\;, \nonumber
\end{eqnarray}
where $\chi = [\gamma_{2-}^2 + (3\gamma_{2+} -\imath \mu)^2 +
  \gamma_{2-} (9\gamma_{2+}- 2\imath \mu)]^{1/2}$. Equation
\eref{eq:psi0_decay_NLD} describes the relaxation of a non-classical state
due to weak thermal excitations. The relaxation rate, which is determined
by $\chi$, depends on the strength of the Kerr constant $\mu$ and
temperature via $\gamma_{2\pm}$. It is interesting to consider the
limiting cases of strong and weak NLD. For strong NLD,
$\gamma_{2-}/\mu\rightarrow \infty$, one finds $\chi \rightarrow
\gamma_{2-}+ (11\gamma_{2+})/2$ and $\psi_0(1,t)$ becomes
\be
\psi_0(1,t) \simeq \psi_0(1,0)e^{-\gamma_{2+} t}.
\ee
Similarly, for weak NLD, $\gamma_{2-}/\mu \rightarrow 0$, 
$\chi \rightarrow -\imath \mu + 4\gamma_{2+} +\gamma_{2-} $
and the time evolution is
\be
\psi_0(1,t)\simeq \psi_0(1,0)e^{(\imath \mu  - 4 \gamma_{2+} )t}.
\ee
Obviously, the decoherence rate increases with increasing strength of the
Kerr constant. This can be understood from the fact that the
accumulated phase shift in the excited state is proportional to
$\mu$. Therefore, for finite $\mu$ there is an additional dephasing,
which contributes to the decay of $\psi_0(1,t)$. Altogether, the
position variance becomes
\be \label{eq:var_finT_NLD}
\mean{\Delta q^2}/q_0^2
\approx\frac{1}{2}+ 2 n_{\rm B}(2\Omega)+ P_{\rm odd} + |\psi_0(1,t)|^2\;.
\ee

\begin{figure}[t]
  \centering
             \includegraphics[width=0.49\textwidth]
                             {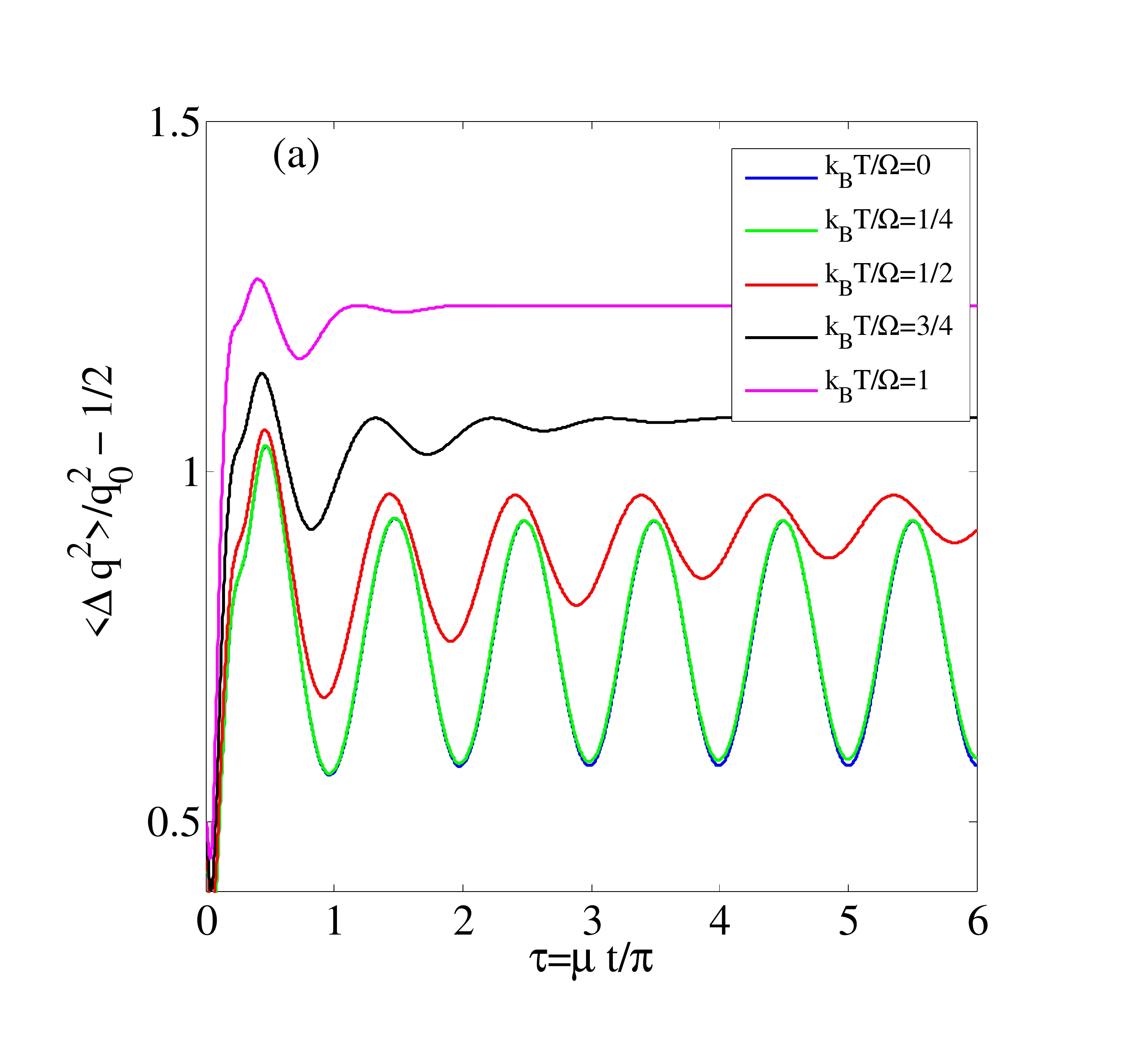}
             \includegraphics[width=0.49\textwidth]
                             {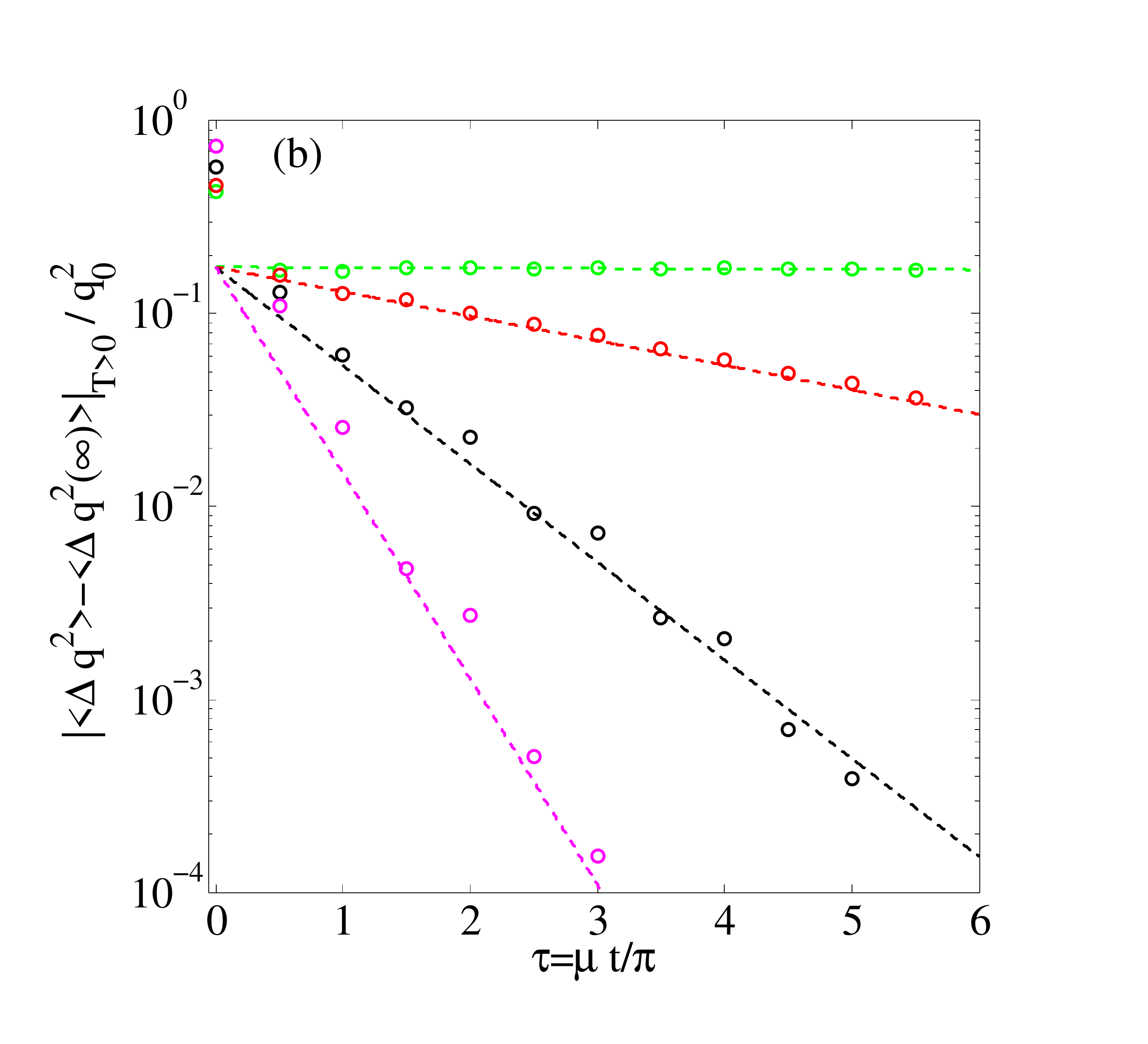} \\      
             \includegraphics[width=0.49\textwidth]
                             {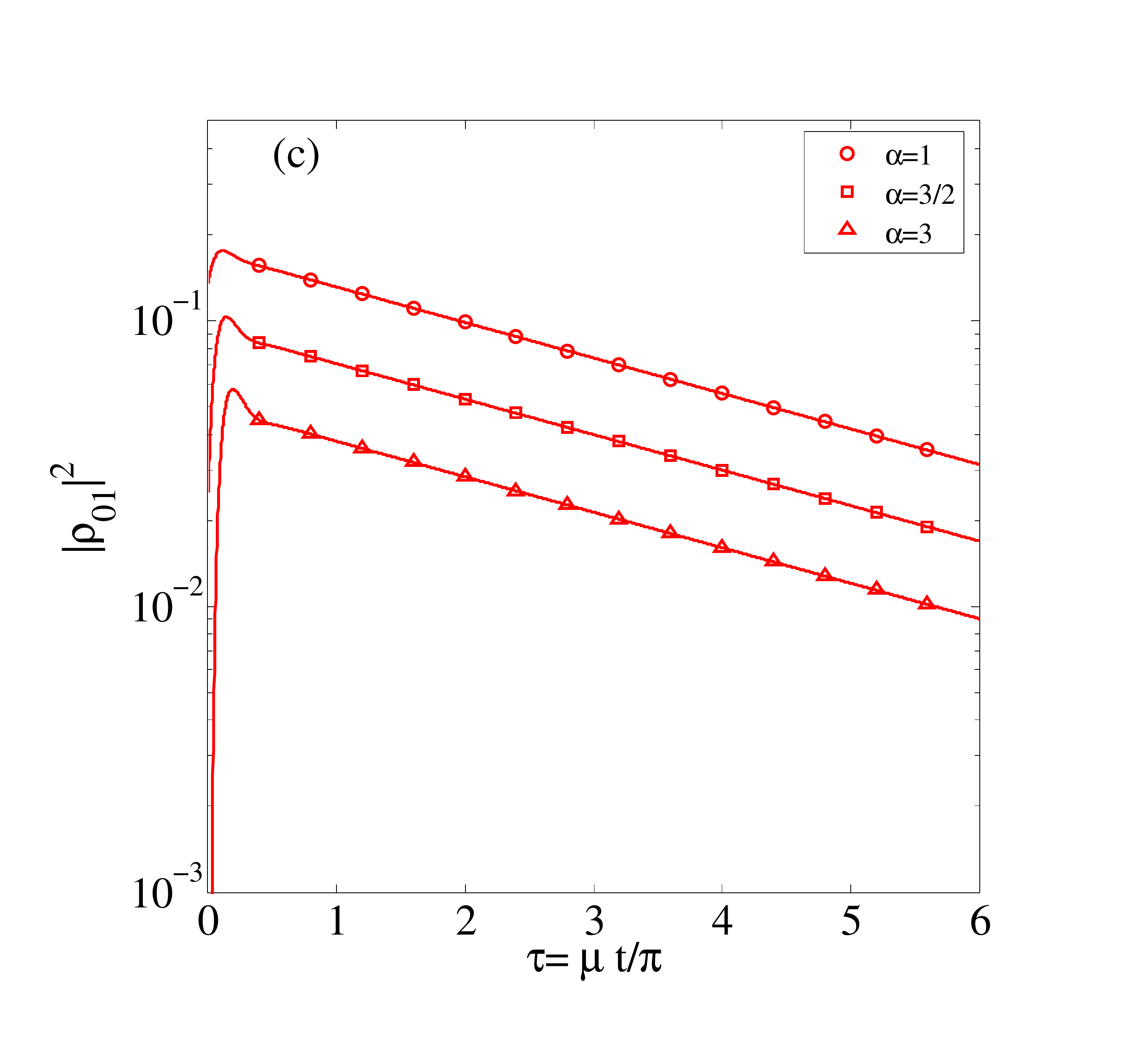}
\caption{ 
(a) Time dependence of the position variance of the nonlinearly damped
  mechanical mode at finite temperature for the same parameters as in
  figure \ref{fig:finT_statvar}.
(b) Time dependence of $\vert \langle \Delta q^2(\tau) \rangle-\langle
 \Delta q^2(\infty) \rangle \vert_{T>0}$ for parameters $\alpha=1$,
 $\mu=\gamma_{2-}$, $\gamma_-=0$ and $k_{\rm B}T/\Omega\in[1/4, 1/2,
   3/4, 1]$. Symbols denote numerical results averaged over one
 period of and dashed lines show the behavior according to
 (\ref{eq:finT_ansatz}).
(c) Time dependence of $\vert \rho_{0,1}(\tau) \vert^2$ at $k_{\rm
   B}T/\Omega=1/2$ for $\alpha=1$ ($\circ$), $\alpha=3/2$ ($\square$)
 and $\alpha=3$ ($\triangle$). The long-time limit relaxation rate
 is according to \eref{eq:sol_psi01}.
  \label{fig:varTfinNLD}}
\end{figure}

Figure \ref{fig:varTfinNLD} (a) shows the time dependence of the
numerically calculated variance for the same parameters as in figure
\ref{fig:stat_state_finT} (d) with the initial displacement amplitude
$\alpha=1$. The $T=0$ variance is included as a reference (blue) and
is almost identical to the curve for $k_{\rm B}T/\Omega=1/4$
(green). When comparing the graphs in figure \ref{fig:varTfinNLD} (a)
with figure \ref{fig:var_relaxT0} (a) and \ref{fig:varT0NLD+LD_alpha1}
(a), one finds a qualitatively similar behavior. It is seen that also
for finite temperatures the relaxation proceeds sequentially. The
short-time dynamics show an increase in variance, corresponding to the
formation of a Schr{\"o}dinger cat state. For longer times the
variance is increasing with time, and the rate of approaching the
stationary value is faster for larger temperatures.
In order to quantify the finite temperature relaxation rate, the time
evolution of $\vert \langle \Delta q^2(\tau) \rangle-\langle \Delta
q^2(\infty) \rangle \vert_{T>0}$ is shown in Figure
\ref{fig:varTfinNLD} (b), where the stationary state variance $\langle
\Delta q^2(\infty) \rangle_{T>0}$ is subtracted from $\langle \Delta
q^2(\tau)\rangle_{T>0}$ in order to remove the finite asymptotic
value. The numerically obtained variance (symbols) is averaged over
one period $t=2\pi/\mu$, which removes the oscillations due to the
finite Kerr constant $\mu$. The temperatures are the same as in figure
\ref{fig:varTfinNLD} (a). The figure confirms the idea of different
relaxation stages, with the decay in the second stage being
exponential and the rate being temperature dependent.

For comparison, figure \ref{fig:varTfinNLD} (c) shows the time
evolution of $\vert \rho_{0,1}(\tau)\vert^2$ for $k_{\rm B}T/\Omega =
1/2$ and $\alpha\in [1,3/2,3]$. It can be seen that in the long-time
limit the decay rates do not depend on the initial oscillator
displacement $\alpha$.
As discussed above, the off-diagonal matrix elements $\rho_{0,1}$ decay
due to thermal excitations. For low temperatures, their evolution is
given by (\ref{eq:psi0_decay_NLD}). When the Kerr constant and the NLD
are equally strong, ($\mu=\gamma_{2-}$), one finds $\chi\rightarrow
\gamma_{2-} + (19\gamma_{2+})/4 -\imath(\mu-3\gamma_{2+}/4)$ and the
time evolution is given by
\be\label{eq:sol_psi01} 
\psi_0(1,t)\simeq \psi_0(1,0)e^{\imath (\mu +
  3\gamma_{2+}/2) t} e^{-(5\gamma_{2+}/2)t}.  
\ee
The resulting behavior is shown in figure \ref{fig:varTfinNLD} (c) and
it can be seen that \eref{eq:sol_psi01} correctly describes the
decay of $\RDM_{0,1}$. From figures \ref{fig:varTfinNLD} (a) and (b) it
can be concluded that the variance increase is mainly due to the decay
of $\vert \rho_{0,1} \vert^2$. Consequently, we make the Ansatz
\be\label{eq:finT_ansatz} 
\vert\langle \Delta q^2(\infty) \rangle_{T>0} -
\langle \Delta q^2(t) \rangle\vert_{T>0} = 
q_0^2 |\psi_0(1,0)|^2 e^{-\Gamma t}, 
\ee
which will be valid in the second stage of the relaxation. From
\eref{eq:sol_psi01} we find $\Gamma(T)\approx 5\gamma_{2+}(2\Omega)$.
This Ansatz is displayed in figure \ref{fig:varTfinNLD} (b) (dashed
lines) and a good quantitative agreement with the numerical results is
found for all temperatures.  Using (\ref{eq:finT_ansatz}) one can
therefore infer the value of $\vert\rho_{0,1} \vert^2$ at the end of
the first relaxation stage, which is dominated by the NLD. This allows
for a reconstruction of the properties of the non-classical state
\eref{eq:qubit_state}, which is created at the end of the first
relaxation stage.

\subsection{Finite temperature, LD and NLD}
\begin{figure}[!ht]
  \centering
             \includegraphics[width=0.49\textwidth]
                             {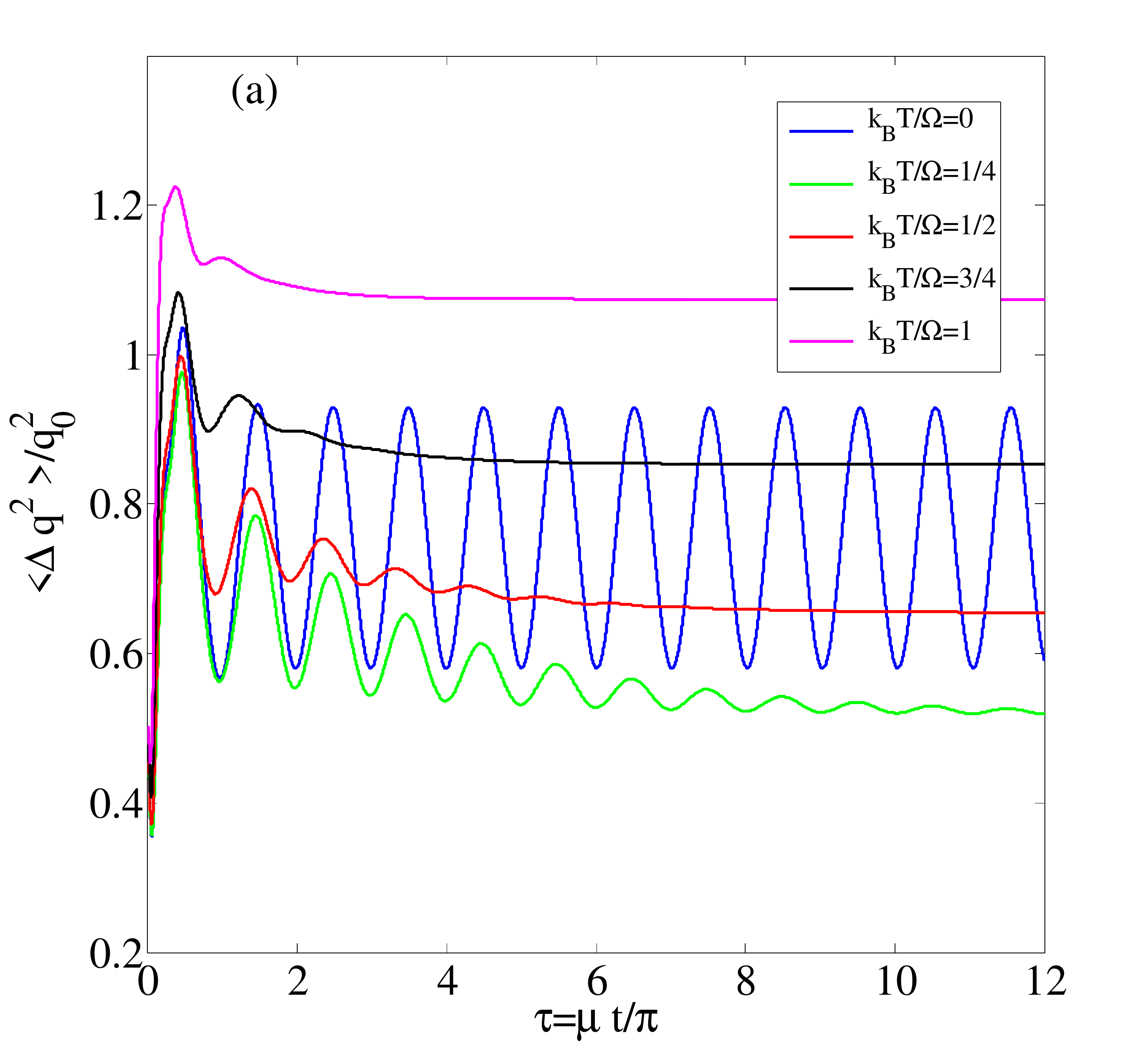}
             \includegraphics[width=0.49\textwidth]
                             {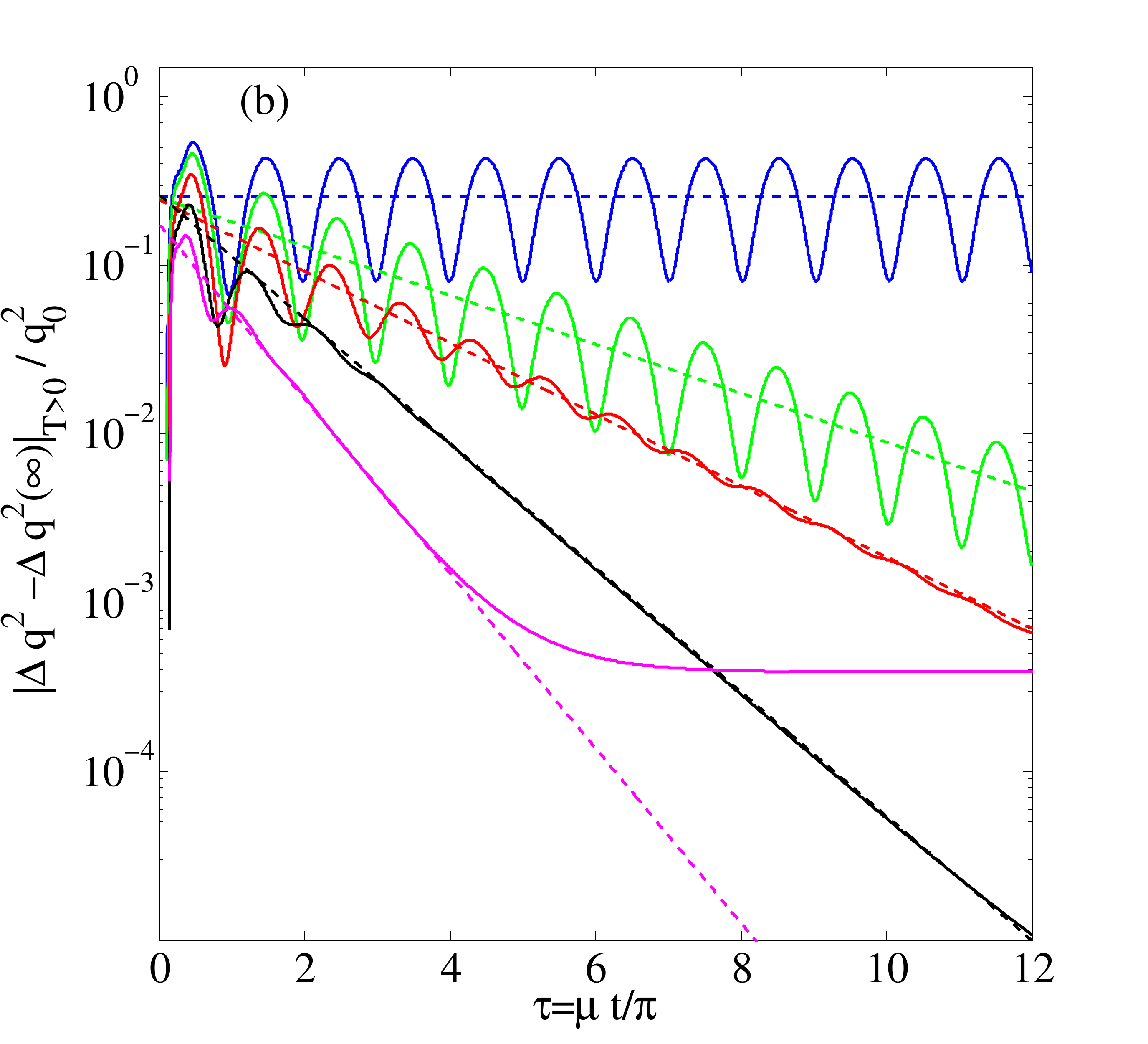}\\
             \includegraphics[width=0.49\textwidth]
                             {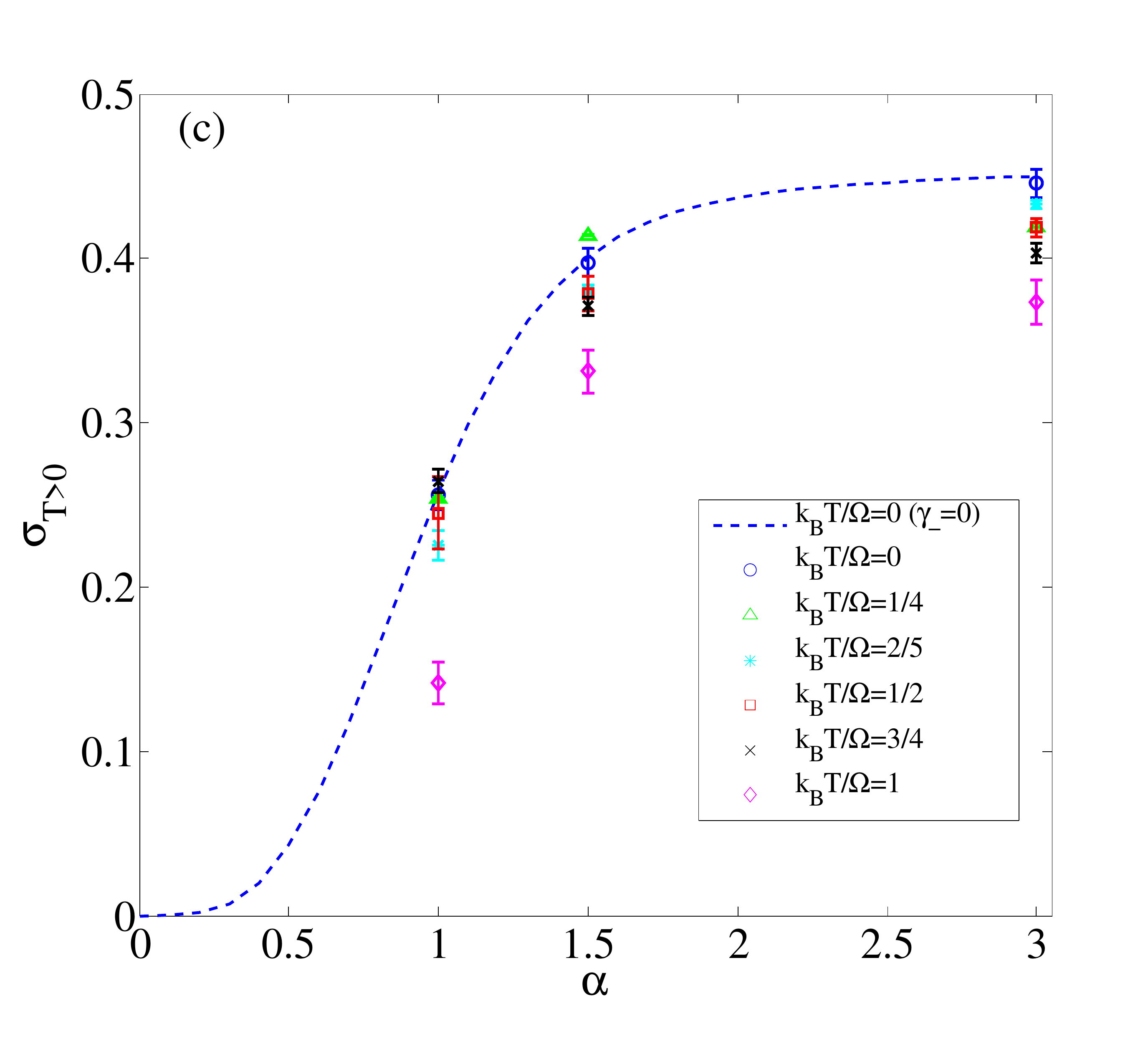}       
             \includegraphics[width=0.49\textwidth]
                             {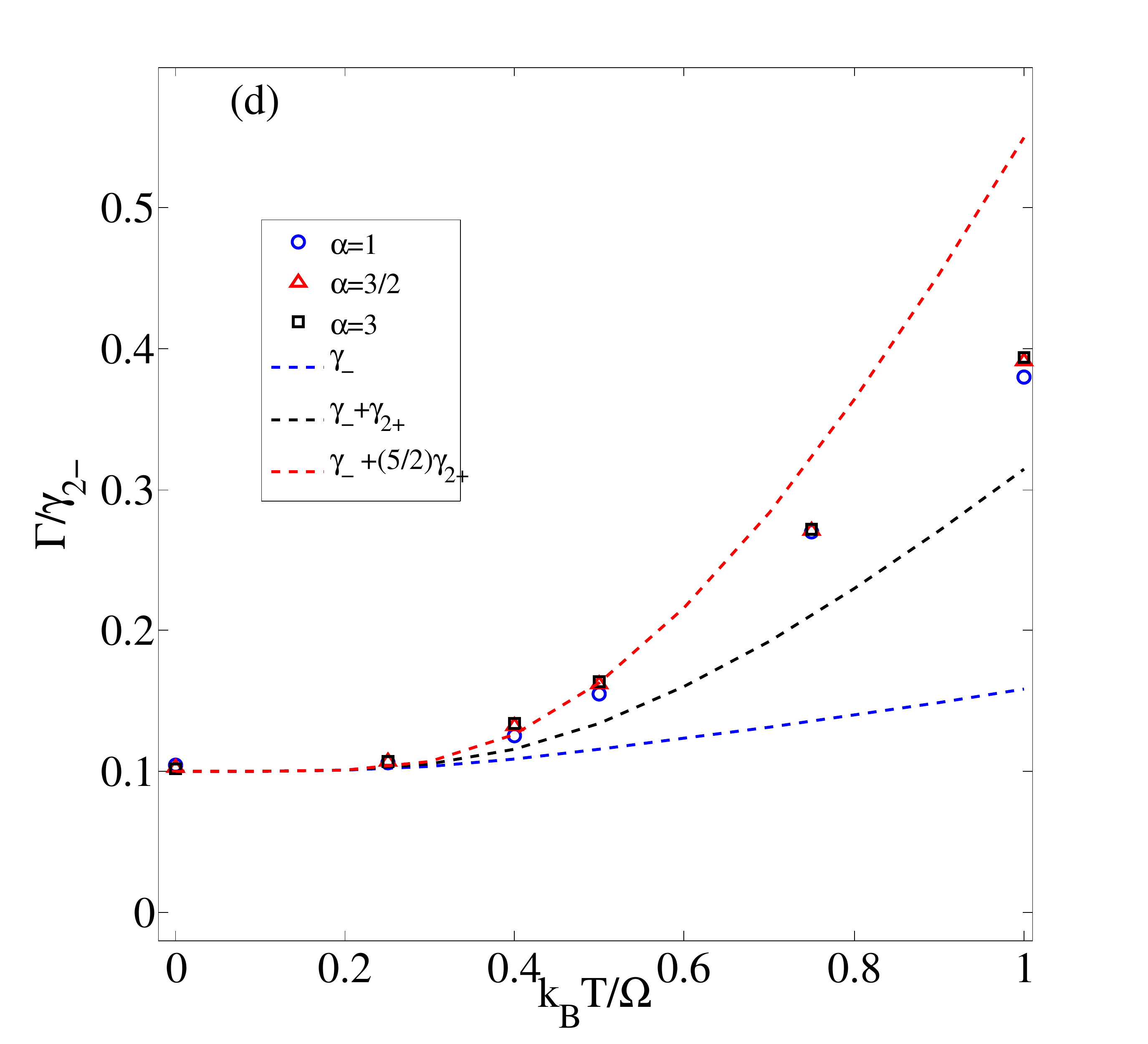}
\caption{(a) Time-dependence of the position variance of an initial
  coherent state with $\alpha=1$, at different temperatures with NLD
  ($\gamma_{2-}=\mu$) and LD ($\gamma_{-}/\gamma_{2-}=1/10$). (b)
  Time dependence of $\vert \langle \Delta q^2(\tau) \rangle-\langle
  \Delta q^2(\infty) \rangle \vert_{T>0}$. The dashed lines indicate
  a fit of \eref{eq:finT_LD_NLD_fit} to the numerical data. (c)
  and (d) Extracted fit parameters $\sigma$ and $\Gamma$ as functions
  of initial displacement amplitude $\alpha$ and temperature,
  respectively. The dashed line in (c) shows the variance according to
  \eref{eq:varT0analytic} for $T=0$ and $\gamma_{-}=0$.
\label{fig:figure8}}
\end{figure}
Finally, in this section we consider the time evolution of the
position variance at finite temperature under influence of both LD and
NLD. Figure \ref{fig:figure8} (a) shows the time-dependence of the
numerically obtained $\mean{\Delta q^2}$ for
$\gamma_{2-}=\mu,~\gamma_{-}/\gamma_{2-}=1/10,~\alpha=1$ and different
temperatures. The behavior for $T=0$, which corresponds to setting
$\gamma_{-}=0$, is included for reference (blue). As in the previous
section we subtract the stationary state variance from $\mean{\Delta
  q^2}$ to study the relaxation towards the stationary state. This is
shown in figure \ref{fig:figure8} (b). For long times an exponential
decay is seen for temperatures $k_{\rm B}T/\Omega <1$. To quantify the
decay we use the Ansatz in \eref{eq:finT_ansatz} and consider
\be\label{eq:finT_LD_NLD_fit}
\vert \langle \Delta q^2 \rangle-\langle \Delta q^2(\infty) \rangle
\vert_{T>0} = q_0^2 (\sigma-\frac{1}{2}) e^{-\Gamma t},
\ee
where $\sigma$ and $\Gamma$ are estimated by a least-square fit to the
numerical data. To this end the variance is averaged over one period
$t=2\pi/\mu$ which removes the oscillations due to the finite Kerr
constant $\mu$. The fitting for $T\leq 3/4$ is done for the time
interval $3<\tau<11$ and for $T=1$ in the time
interval of $2<\tau<4$. The resulting behavior of
\eref{eq:finT_LD_NLD_fit} is displayed as dashed lines in figure
\ref{fig:figure8} (b).

In figures \ref{fig:figure8} (c) and (d) the extracted parameters
$\sigma$ and $\Gamma$ are shown as functions of initial displacement
amplitude $\alpha$ and temperature, respectively. Due to the
sequential relaxation, the parameter $\sigma$ approximately
corresponds to the variance of the non-classical state
\eref{eq:qubit_state}, which is created in the initial stage.  The
analytic expression for $T=0$, given by equation
\eref{eq:varT0analytic}, is shown in figure \ref{fig:figure8} (c) for
comparison. The extracted low temperature results are in good
agreement with the analytic curve. For higher temperatures deviations
from the analytic behavior can be seen. However, the qualitative
dependence on the initial amplitude is still similar to the one given
by \eref{eq:varT0analytic}.

In figure \ref{fig:figure8} (d) the temperature dependence of the
extracted decay rates $\Gamma$ is compared to the rates of one-vibron
loss $\gamma_-$ and two-vibron excitations $\gamma_{2+}$, which
determine the decay rate in the situations investigated in sections
\ref{sec:res_finT_NLD} and \ref{sec:res_zeroT_LD_NLD}. It can be seen,
that also in the present case the decay rate is independent of the
initial coherent state displacement amplitude. For low temperatures,
$k_{\rm B}T/\Omega<0.3$, one-vibron loss is the dominating damping
mechanism. For higher temperatures the interplay of one-vibron and
two-vibron processes leads to a more complicated behavior. At first
the rate follows the behavior given by the combined rates
$\gamma_{-}+5\gamma_{2+}/2$ but starts to deviate at $k_{\rm
  B}T/\Omega\approx0.3$. With increasing temperature it stays between
$\gamma_{-}+\gamma_{2+}$ and $\gamma_{-}+5\gamma_{2+}/2$.

The results presented in this section show that a ring-down
measurement of the variance can be used to identify the quantum features
of an initially created non-classical state, provided the sequential
relaxation regime can be accessed. Moreover, the analysis of the decay
rates can provide information about the dominating
dissipation mechanisms in the system under investigation.

\section{Summary and conclusions}\label{sec:summary}
In this work we have investigated a nonlinear oscillator in the
quantum regime. The system is subject to nonlinear damping (two-vibron
exchange), which resembles the two-photon decay known from quantum
optics. The initially displaced oscillator relaxes to a steady
state, which is determined by the relative strength of the linear and
the nonlinear damping mechanisms and the temperature of the thermal
bath. For sufficiently strong NLD we find that the relaxation is
sequential: In the first stage, the initial density matrix decays to a
non-classical state \eref{eq:qubit_state}, followed by the second
stage, where this state further decays to the final steady
state. The non-classical state depends on the initial state, while the
decay rates depend on temperature and the strength of LD and NLD.

At zero temperature, the sequential relaxation allows for a
reconstruction of the properties of the non-classical state even in
the presence of LD, i.e., when the stationary state is always the
ground state. This can be achieved by performing a ring-down
measurement of the position variance. For finite temperatures, NLD and
thermal excitations lead to a different steady state - a
stationary thermal state with a parity dependent Boltzmann-like
distribution \eref{eq:finT_NLD}. The parity dependence is a signature
of the NLD at finite temperatures. In the sequential relaxation regime
the properties of the non-classical state can again be extracted from
variance measurements.
 
In the following we briefly discuss the possibility of observing the
relaxation of the non-classical state \eref{eq:qubit_state} in
experiments with carbon-based nanomechanical resonators. A recent
experiment \cite{eimo+11} indicates significant NLD in micrometer
sized, carbon-based resonators (nanotubes and graphene) with resonance
frequencies of several hundreds of MHz. These systems are
nonlinear and can be described by the classical Duffing equation of
motion \cite{licr08}
\be\label{eq:duffing}
m\ddot{q}= - m \Omega^2 q  -\alpha_0 q^3 -\gamma \dot{q}-
\eta q^2\dot{q},
\ee
where $q$ is the oscillator position, $m$ and $\Omega$ are the
oscillator mass and frequency, $\alpha_0$ is the nonlinearity
strength, $\gamma$ is the LD rate and $\eta$ is the NLD rate. By
comparing \eref{eq:duffing} with the corresponding quantum mechanical
equation, the relation between the parameters ($\mu, \gamma_{-},
\gamma_{2-}$) and the parameters of the Duffing equation can
be established,
\begin{eqnarray}
\mu = \frac{3\hbar\alpha_0}{8m^2\Omega^3}\:\:\:\: & \gamma_{2-}=\frac{\hbar
  \eta}{4m^2\Omega^2}\:\:\:\:&
\gamma_{-}=\frac{\gamma}{m\Omega}.
\end{eqnarray}
The temperature for reaching the quantum regime without cooling is of
the order of $T\lesssim 10$ mK for resonators with frequencies in the
$100$ MHz range.  The values obtained in \cite{eimo+11} correspond to
$\mu/\Omega\ll 10^{-3}$, meaning that the RWA-analysis works well. The
magnitude of the nonlinear damping rate is found to yield
$\gamma_{2-}\approx \mu$, which is promising for the formation of a
non-classical state. The condition to find the sequential relaxation
regime amounts to
\begin{eqnarray}\label{eq:rg_cond}
\gamma_{-}<\gamma_{2-}
\longleftrightarrow\gamma<
\frac{\hbar\eta}{4m\Omega}= \frac{1}{4}q_0^2\eta,
\end{eqnarray}
where $q_0$ is the zero point amplitude. Consequently, for systems
with a large zero point amplitude it is easier to reach the
sequential relaxation regime. In this regard, carbon-based resonators
are very promising due to their low mass and tunable frequency. In
present experiments the condition $\gamma_{-}<\gamma_{2-}$ is not yet
fulfilled, but the ongoing experimental and technological progress in
the field will help to improve the parameters in order to meet the
condition (\ref{eq:rg_cond}).

In conclusion, our study shows that exploiting NLD present in
(carbon-based) NEMS resonators is a promising route to generation and
investigation of non-classical effects in NEMS. Studies of the
sequential relaxation in multi-partite systems, such as coupled
oscillators, would be of interest with respect to the generation of
quantum entanglement.
\ack We thank J. M.~Kinaret and L. Y.~Gorelik for helpful
discussion. The research leading to this article has received funding
from the Swedish Research Council.

\appendix 
\section{Derivation of equations \eref{eq:C2n} and \eref{eq:psi_t_even}.}\label{sec:app_zeroT_NLD}
For zero temperature we set $\gamma_{2+}=0$ and get
\begin{eqnarray}
  \frac{\partial}{\partial t} \rho_{n,n}(t) &=& \gamma_{2-} 
  (n+2)(n+1) \rho_{n+2,n+2} - \gamma_{2-} n(n-1)\rho_{n,n}\;,\\
  \frac{\partial}{\partial t} \psi_n(1,t) &=& \gamma_{2-} (n+2)(n+1) \psi_{n+2} 
  -\left[ -\imath \mu (1+2n) + \gamma_{2-} n^2 \right] \psi_n\;.
\end{eqnarray}
One sees that for $t\to\infty$ only $\rho_{0,0}$, $\rho_{1,1}$ and
$\psi_0(1)$ remain non-zero (depending on initial conditions). To find
the respective values in the long time limit, one first verifies that
\begin{equation}
  \frac{\partial}{\partial t} \sum_n \rho_{2n,2n} = 0\;,\quad
  \frac{\partial}{\partial t} \sum_n \rho_{2n+1,2n+1} = 0\;.
\end{equation}
Therefore, $\rho_{0,0}(\infty) = \sum_n \rho_{2n,2n}(0)$ and
$\rho_{1,1}(\infty) = \sum_n \rho_{2n+1,2n+1}(0)$ are constants of
motion. There exists another constant of the motion, which can be
found as follows: We take $\psi_n = e^{\imath \mu t} \tilde{\psi}_n$
and consider
\begin{equation}
\frac{\partial}{\partial t} \sum_n C_{2n} \tilde{\psi}_{2n} =
\gamma_{2-}\sum_{n=0} C_{2n} \left( (2n+2)(2n+1) \tilde{\psi}_{2n+2}
-\left[ -\imath \frac{\mu}{\gamma_{2-}} 4n + 4n^2 \right]
\tilde{\psi}_{2n} \right)
\end{equation}
with so far unknown coefficients $C_{2n}$. Those shall be determined
such that the right hand side vanishes. This is the case if
\begin{equation}
C_{2n-2} (2n-1) = \left[ -\imath 2\frac{\mu}{\gamma_{2-}} + 2n \right]
C_{2n}
\end{equation}
is fulfilled. This recursion can be solved in terms of Gamma functions
\begin{equation}
  C_{2n} = \frac{\Gamma(\frac{1}{2} + n) 
    \Gamma(1-\imath \frac{\mu}{\gamma_{2-}})}{\sqrt{\pi}
    \Gamma(1 + n-\imath \frac{\mu}{\gamma_{2-}}) }\;,
\end{equation}
which can be verified using the identity $\Gamma(z+1) = z \Gamma(z)$.
Hence, $\psi_0(1, \infty) = \sum_n C_{2n}
\tilde{\psi}_{2n}(0)$.

If the initial state is a coherent state,
\begin{equation}
\RDM_{n,m} (0) = \exp(-|\alpha|^2) \alpha^n \alpha^{*m}/\sqrt{n!m!}\;,
\end{equation}
one gets
\be
\sum_n C_{2n} \tilde{\psi}_{2n}(0)
= \exp(-|\alpha|^2) \alpha^* \sum_n C_{2n} \frac{z^{n}}{(2n)!}
\equiv \exp(-|\alpha|^2) \alpha^* F(z)\;
\ee
with $z=|\alpha|^4$. It can be shown, that $F(z)$ corresponds to the
definition of the confluent hypergeometric limit function ${}_0F_1$
\cite{AS65}. Consequently,
\begin{equation}
  \tilde{\psi}_{0}(1,\infty) \equiv \sum_n C_{2n}
  \tilde{\psi}_{2n}(0) = \exp(-|\alpha|^2) \alpha^*
        {}_0F_1(1-\imath\frac{\mu}{\gamma_{2-}},|\alpha|^4/4)\;.
\end{equation}

\section{Derivation of equations \eref{eq:finT_NLD} and \eref{eq:finT_n}}\label{sec:app_fin_temp}
The equations for finite temperature can be directly obtained from
\eref{eq:nonlinQMEPsi}:
\be
\begin{array}{lll}
\frac{\partial}{\partial t} \rho_{n,n}(t)& =&
\gamma_{2+}n(n-1)\rho_{n-2,n-2}+ \gamma_{2-} (n+2)(n+1) \rho_{n+2,n+2}\\ 
& & -\left[ \gamma_{2-} n(n-1) + \gamma_{2+}
(n+1)(n+2)\right]\rho_{n,n}\;,\\ 
\frac{\partial}{\partial t}
\psi_n(1,t)& =& \gamma_{2+}n(n+1)\psi_{n-2}+\gamma_{2-} (n+2)(n+1)
\psi_{n+2} \\ 
\nonumber & & - \left[ -\imath \mu (1+2n) + \gamma_{2-}
 n^2 + \gamma_{2+}(n+2)^2 \right] \psi_n\;.
\end{array}
\ee
The stationary probability distribution $\rho_{n,n}(\infty)$ obeys
the detailed balance condition
\begin{equation}
  \gamma_{2+} \rho_{n,n}(\infty) = \gamma_{2-} \rho_{n+2,n+2}(\infty)\;.
\end{equation}
Therefore,
\begin{eqnarray}
\rho_{2n,2n}(\infty) = \left(\frac{\gamma_{2+}}{\gamma_{2-}}\right)^n
\rho_{0,0}(\infty) = e^{-2 \hbar \Omega \beta
  n}\rho_{0,0}(\infty)\;,\\ \nonumber \rho_{2n+1,2n+1}(\infty) =
\left(\frac{\gamma_{2+}}{\gamma_{2-}}\right)^n \rho_{1,1}(\infty) =
e^{-2\hbar \Omega \beta n} \rho_{1,1}(\infty)\;.
\end{eqnarray}
Since parity is conserved one can use
\begin{eqnarray}
\sum_n \rho_{2n,2n}(\infty) = \rho_{0,0}(\infty) \sum_n
e^{-2\hbar\Omega \beta n} = P_{\rm even}\;, \\\nonumber \sum_n
\rho_{2n+1,2n+1}(\infty) = \rho_{1,1}(\infty) \sum_n e^{-2\hbar \Omega
  \beta n} = P_{\rm odd}\;
\end{eqnarray}
to fix the values $\rho_{0,0}$ and $\rho_{1,1}$. The sum evaluates to
$\sum_n e^{-2\hbar \Omega \beta n} = 1 + n_B(2\Omega)$. With these
results it is straight-forward to calculate the average number of
excitations
\begin{eqnarray*}
\sum_{n} 2n \rho_{2n,2n}(\infty) &=& \frac{2 P_{\rm even}}{1 +
  n_B(2\Omega)} \sum_{2n} n e^{-2\hbar \Omega \beta n}\\ & =& \frac{2
  P_{\rm even}}{1 + n_B(2\Omega)} n_B(2\Omega)^2 \Bigg(
\frac{1}{n_B(2\Omega)}+ 1\Bigg) = 2 P_{\rm even}
n_B(2\Omega)\\ \sum_{n} (2n+1) \rho_{2n+1,2n+1}(\infty) &=& \frac{2
  P_{\rm odd}}{1 + n_B(2\Omega)} \sum_{2n} n e^{-2\hbar \Omega \beta
  n} + P_{\rm odd} = P_{\rm odd}(2 n_B(2\Omega) + 1).
\end{eqnarray*}
And therefore
\begin{equation}
\mean{a^{\dag}a} = \sum_{2n} 2n \rho_{2n,2n}(\infty) + \sum_{2n} (2n+1)
\rho_{2n+1,2n+1}(\infty) = 2 n_B(2\Omega) + P_{\rm odd}\;.
\end{equation}
%

\bibliographystyle{unsrt}
\bibliography{mprnmo}

\begin{thebibliography}{10}

\bibitem{poza12}
Menno Poot and Herre S.~J. van~der Zant.
\newblock Mechanical systems in the quantum regime.
\newblock {\em Physics Reports}, 511:273--335, 2012.

\bibitem{ocho+10}
A.~D. O'Connell, M.~Hofheinz, M.~Ansmann, Radoslaw~C. Bialczak, M.~Lenander,
  Erik Lucero, M.~Neeley, D.~Sank, H.~Wang, M.~Weides, J.~Wenner, John~M.
  Martinis, and A.~N. Cleland.
\newblock Quantum ground state and single-phonon control of a mechanical
  resonator.
\newblock {\em Nature}, 464:697--703, 2010.

\bibitem{tedo+11}
J.~D. Teufel, T.~Donner, Dale Li, J.~W. Harlow, M.~S. Allman, K.~Cicak, A.~J.
  Sirois, J.~D. Whittaker, K.~W. Lehnert, and R.~W. Simmonds.
\newblock Sideband cooling of micromechanical motion to the quantum ground
  state.
\newblock {\em Nature}, 475:359--363, 2011.

\bibitem{chal+11}
J.~Chan, T.P.M. Alegre, A.H. Safavi-Naeini, J.T. Hill, A.~Krause,
  S.~Gr{\"o}blacher, M.~Aspelmeyer, and O.~Painter.
\newblock Laser cooling of a nanomechanical oscillator into its quantum ground
  state.
\newblock {\em Nature}, 478(7367):89--92, 2011.

\bibitem{voki+12}
A.~Voje, J.~M. Kinaret, and A.~Isacsson.
\newblock Generating macroscopic superposition states in nanomechanical
  graphene resonators.
\newblock {\em Phys. Rev. B}, 85:205415, 2012.

\bibitem{riki+12}
S~Rips, M~Kiffner, I~Wilson-Rae, and M~J Hartmann.
\newblock Steady-state negative {W}igner functions of nonlinear nanomechanical
  oscillators.
\newblock {\em New Journal of Physics}, 14(2):023042, 2012.

\bibitem{atis+08}
J.~Atalaya, A.~Isacsson, and J.~M. Kinaret.
\newblock Continuum elastic modeling of graphene resonators.
\newblock {\em Nano Lett.}, 8(12):4196--4200, 2008.

\bibitem{zw73}
R.~Zwanzig.
\newblock Nonlinear generalized {L}angevin equations.
\newblock {\em J. Stat. Phys.}, 9:215--220, 1973.

\bibitem{lise81}
K.~Lindenberg and V.~Seshadri.
\newblock Dissipative contributions of internal multiplicative noise: {I}.\
  mechanical oscillator.
\newblock {\em Physica A}, 109:483--499, 1981.

\bibitem{dykr84}
M.~Dykman and M.~Krivoglaz.
\newblock Theory of nonlinear oscillator interacting with a medium.
\newblock {\em Soviet Scientific Reviews, Section A, Physics Reviews},
  5:265--441, 1984.

\bibitem{zash+12}
S.~Zaitsev, O.~Shtempluck, E.~Buks, and O.~Gottlieb.
\newblock Nonlinear damping in a micromechanical oscillator.
\newblock {\em Nonlinear Dynam.}, 67:859--883, 2012.

\bibitem{eimo+11}
A.~Eichler, J.~Moser, J.~Chaste, M.~Zdrojek, I.~Wilson-Rae, and A.~Bachtold.
\newblock Nonlinear damping in mechanical resonators made from carbon nanotubes
  and graphene.
\newblock {\em Nat. Nanotechnol.}, 6:339--342, 2011.

\bibitem{crmi+12}
A.~Croy, Midtvedt D., A.~Isacsson, and J.~M. Kinaret.
\newblock Nonlinear damping in graphene resonators.
\newblock {\em Phys. Rev. B}, 86:235435, 2012.

\bibitem{silo78}
H.~D. Simaan and R.~Loudon.
\newblock Off-diagonal density matrix for single-beam two-photon absorbed
  light.
\newblock {\em Journal of {P}hysics {A}: {M}athematical and {G}eneral}, 11:435,
  1978.

\bibitem{gikn93}
L.~Gilles and P.~L. Knight.
\newblock Two-photon absorption and nonclassical states of light.
\newblock {\em Phys. Rev. A}, 48:1582--1593, 1993.

\bibitem{giga+94}
L.~Gilles, B.~M. Garraway, and P.~L. Knight.
\newblock Generation of nonclassical light by dissipative two-photon processes.
\newblock {\em Phys. Rev. A}, 49:2785--2799, 1994.

\bibitem{lo84}
R~Loudon.
\newblock Squeezing in resonance fluorescence.
\newblock {\em Opt. Commun.}, 49:24--28, 1984.

\bibitem{evsp+12}
M.~J. Everitt, T.~P. Spiller, G.~J. Milburn, R.~D. Wilson, and A.~M. Zagoskin.
\newblock Cool for cats.
\newblock arXiv:1212.4795, 2012.

\bibitem{sh67}
Y.~R. Shen.
\newblock Quantum statistics of nonlinear optics.
\newblock {\em Phys. Rev.}, 155:921--931, 1967.

\bibitem{nubo+10}
A.~Nunnenkamp, K.~B\o{}rkje, J.~G.~E. Harris, and S.~M. Girvin.
\newblock Cooling and squeezing via quadratic optomechanical coupling.
\newblock {\em Phys. Rev. A}, 82:021806, 2010.

\bibitem{wabl53}
R.~K. Wangsness and F.~Bloch.
\newblock The dynamical theory of nuclear induction.
\newblock {\em Phys. Rev.}, 89(4):728--739, 1953.

\bibitem{bl57}
F.~Bloch.
\newblock Generalized theory of relaxation.
\newblock {\em Phys. Rev.}, 105(4):1206--1222, 1957.

\bibitem{re65}
A.~G. Redfield.
\newblock The theory of relaxation processes.
\newblock {\em Adv. Magn. Reson.}, 1:1, 1965.

\bibitem{gl63}
Roy~J. Glauber.
\newblock Coherent and incoherent states of the radiation field.
\newblock {\em Phys. Rev.}, 131:2766--2788, 1963.

\bibitem{yust86}
B.~Yurke and D.~Stoler.
\newblock Generating quantum mechanical superpositions of macroscopically
  distinguishable states via amplitude dispersion.
\newblock {\em Phys. Rev. Lett.}, 57(1):13--16, 1986.

\bibitem{brpe02}
H.~P. Breuer and F.~Petruccione.
\newblock {\em The Theory of Open Quantum Systems}.
\newblock Oxford University Press, USA, 2002.

\bibitem{dy12}
M.~Dykman.
\newblock Private communication.

\bibitem{gazo04}
C.~W. Gardiner and P.~Zoller.
\newblock {\em Quantum Noise: {A} Handbook of {M}arkovian and Non-{M}arkovian
  Quantum Stochastic Methods with Applications to Quantum Optics}.
\newblock Springer, 2004.

\bibitem{miho86}
G.~J. Milburn and C.~A. Holmes.
\newblock Dissipative quantum and classical {L}iouville mechanics of the
  anharmonic oscillator.
\newblock {\em Phys. Rev. Lett.}, 56:2237--2240, 1986.

\bibitem{ph90}
Simon J.~D. Phoenix.
\newblock Wave-packet evolution in the damped oscillator.
\newblock {\em Phys. Rev. A}, 41:5132--5138, 1990.

\bibitem{mo06}
H\'{e}ctor Moya-Cessa.
\newblock Decoherence in atom-field interactions: A treatment using
  superoperator techniques.
\newblock {\em Physics Reports}, 432(1):1 -- 41, 2006.

\bibitem{drwa79}
D.~F.~Walls P.~D.~Drummond.
\newblock Quantum theory of optical bistability. i: Nonlinear polarisability
  model.
\newblock {\em J. Phys. A}, 13:725--741, 1979.

\bibitem{la62}
R.~Landauer.
\newblock Fluctuations in bistable tunnel diode circuits.
\newblock {\em Journal of Applied Physics}, 33:2209--2216, 1962.

\bibitem{ha74}
H.~Haken.
\newblock Exact stationary solution of the master equation for systems far from
  thermal equilibrium in detailed balance.
\newblock {\em Physics Letters A}, 46:443--444, 1974.

\bibitem{haha79}
G.~Haag and P.~H\"{a}nggi.
\newblock Exact solutions of discrete master equations in terms of continued
  fractions.
\newblock {\em Z. Physik B}, 34:411--417, 1979.

\bibitem{domi97}
V.~V. Dodonov and S.~S. Mizrahi.
\newblock Exact stationary photon distributions due to competition between one-
  and two-photon absorption and emission.
\newblock {\em Journal of {P}hysics {A}: {M}athematical and {G}eneral},
  30(16):5657, 1997.

\bibitem{we03}
Eric~W. Weisstein.
\newblock {\em CRC Concise Encyclopedia of Mathematics}.
\newblock Chapman and Hall, 2003.

\bibitem{sc01}
W.~P. Schleich.
\newblock {\em Quantum Optics in Phase Space}.
\newblock Wiley-VCH, Berlin, 2001.

\bibitem{licr08}
R.~Lifshitz and M.C. Cross.
\newblock {\em Nonlinear Dynamics of Nanomechanical and Micromechanical
  Resonators}, chapter~1.
\newblock Wiley-VCH, 2008.

\bibitem{AS65}
M.~Abramowitz and I.A. Stegun.
\newblock {\em Handbook of Mathematical Functions}.
\newblock Dover Pub., 1965.

\end{thebibliography}
\end{document}